\documentclass[12pt]{iopart}
\usepackage{graphicx}

\expandafter\let\csname equation*\endcsname\relax
\expandafter\let\csname endequation*\endcsname\relax

\usepackage{amsmath}
\usepackage{xcolor}
\usepackage{siunitx}

\newcommand{\ket}[1]{\big|#1\big>}

\usepackage{iopams}  
\begin{document}

\title[Photonic analogues of chiral symmetric topological tight-binding models]{Asymptotically exact photonic analogues of chiral symmetric topological tight-binding models}

\author{S. Palmer$^1$, Y. Ignatov$^1$, R.V. Craster$^2$ \& M. Makwana$^2$}

\address{$^1$ The Blackett Laboratory, Imperial College London, London SW7 2AZ, United Kingdom}
\address{$^2$ Department of Mathematics, Imperial College London, London, SW7 2AZ, United Kingdom}
\vspace{10pt}
\begin{indented}
\item[]\today
\end{indented}

\begin{abstract}
Topological photonic edge states, protected by chiral symmetry, are attractive for guiding wave energy as they can allow for more robust guiding and greater control of light than alternatives; however, for photonics, chiral symmetry is often broken by long-range interactions. We look to overcome this difficulty by exploiting the topology of networks, consisting of voids and narrow connecting channels, formed by the spaces between closely spaced perfect conductors. In the limit of low frequencies and narrow channels, these void-channel systems have a direct mapping to analogous discrete mass-spring systems in an asymptotically rigorous manner and therefore only have short-range interactions. We demonstrate that the photonic analogues of topological tight-binding models that are protected by chiral symmetries, such as the SSH model and square-root semimetals, are reproduced for these  void-channel networks with appropriate boundary conditions. We anticipate, moving forward, that this paper provides a basis from which to explore continuum photonic topological systems, in an asymptotically exact manner, through the lens of a simplified tight-binding model.
\end{abstract}

%
%
%
%
%

\section{Introduction}\label{sec:introduction}



The field of topological materials has revealed exotic phenomena such as robust, unidirectional edge states that occur at the interfaces between materials that belong to two different topological phases. The Nobel Prize in Physics 2016 was awarded to Thouless, Haldane, and Kosterlitz, \cite{kosterlitz1973ordering,haldane1983continuum,haldane1983nonlinear} for predicting such phases in electronic systems where such topological edge states promise to revolutionise electronics and quantum computing \cite{checkelsky2012dirac,chiu2016classification,burkov2010spin,vali2015scheme,he2019topological}. Although topological phases were first discovered in electronic systems, the underlying principles are applicable to wave systems in general, including photonic and acoustic systems \cite{zhang2018topological,ozawa2019topological}. There is now great interest in reproducing topological phases in photonics using \emph{photonic crystals}: periodic nanostructures with tunable photonic bands. In the short to medium term, topological photonic materials may improve the performance of photonic devices by reducing dissipation, especially when guiding light around sharp corners, and in the longer term could offer a platform for fault-tolerant quantum computers in photonics \cite{ozawa2019topological,von2020fourty,kim2020recent}. 
Applications that are unique to topological photonic materials include the design of more efficient lasers, where the topological edge mode acts as a cavity in which light propagates and is amplified unidirectionally and coherently despite imperfections in the crystal \cite{rider2019perspective,bandres2018topological,ota2018topological}, and the cloaking of large photon sources from each other using the polarisation of light \cite{khanikaev2013photonic}.

Many topological materials are protected by chiral symmetry (also known as sublattice symmetry) \cite{schnyder2008classification,kitaev2009periodic,ryu2010topological}, both directly as in the SSH model \cite{asboth2016short} and indirectly as in \emph{square-root topological insulators} \cite{arkinstall2017topological} where the squared Hamiltonian is block-diagonal and at least one of the blocks corresponds to a known non-trivial system. Chiral symmetry acts on Bloch Hamiltonians as \cite{chiu2016classification}
\begin{equation}
    \hat{\mathcal{S}}\hat{H}(\vec{k})\hat{\mathcal{S}}^{-1}=-\hat{H}(\vec{k}),
\end{equation}
where $\hat{\mathcal{S}}$ is a unitary operator that squares to $+1$.
Chiral symmetry is relatively common in tight-binding models of electronic systems, however in photonics, chiral symmetry is often broken by long-range interactions \cite{pocock2019bulk,pocock2020thesis}; despite this, it can be engineered in certain photonic systems, examples include arrays of dielectric resonators \cite{poli2015selective} or grating structures \cite{malkova2009observation}. 

In this paper, we engineer chiral symmetric photonic systems where transverse-electric polarised light propagates, in the voids and narrow connecting channels, located between perfect conductors, as shown in figure \ref{fig:void-channel-to-mass-spring}(a). In section \ref{sec:methodology}, we review why at low frequencies, this photonic system behaves like a \emph{discrete} classical network of inductors and capacitors (or equivalently as a classical network of masses and springs, as shown in figure \ref{fig:void-channel-to-mass-spring}(b)), in the limit where the inclusions are closely
spaced \cite{vanel2017asymptotic,vanel2018asymptotic}. The void-channel networks have vanishing long-range interactions, as the channels are made increasingly narrow, thus certain configurations of the void-channel networks have a chiral symmetric limit (up to a shift of constant frequency).

In tight-binding models, terminating the lattice freely does not change the onsite potential and therefore preserves chiral symmetry. In mass-spring and void-channel models, the free boundary condition generally breaks chiral symmetry \cite{maimaiti2020microwave,wakao2020topological}. We propose that the chiral symmetry can be restored at the interfaces by ``capping'' the mass-spring/void-channel networks with heavy masses/large voids, respectively. In section \ref{sec:SSH} we use the well known SSH model \cite{asboth2016short} to demonstrate this principle, and verify that the void-channel SSH geometry features topological edge states. Although in this article we do not concentrate upon the asymptotic theory of mapping continuum systems to discrete models we note that there is considerable advantage in being able to accurately map between them: the entire machinery and theory for topological tight-binding systems then carries across into continuum systems. In simpler settings of connected  acoustic tubes and straight channels  \cite{Zheng_2019,Zheng_2020} and \cite{ZHENG2021100299} illustrate the power of being able to translate back-and-forth between continuum and discrete networks;  we use the asymptotic methodology of  \cite{vanel2017asymptotic,vanel2018asymptotic} showing that curved thin channels can be employed for closely spaced cylinders (and other smooth objects) and noting that a three-dimensional network extension \cite{vanel2019asymptotic} is also available.  

The void-channel geometries are also a promising platform to realise square-root topological systems \cite{arkinstall2017topological}. For example, Maimaiti \emph{et al} \cite{maimaiti2020microwave} showed that the nearest-neighbour tight-binding model of the honeycomb-kagome lattice is a \emph{square-root semimetal} that inherits the topology from the honeycomb sector of the squared Hamiltonian. The authors also proposed an analogous mass-spring model using a gravitational potential energy term to adjust the onsite terms of the equations of motion \cite{maimaiti2020microwave};  it is not apparent to us if an analogue of this gravitational term exists for the void-channel geometries. In section \ref{sec:honeycomb-kagome}, we produce a photonic analogue of the square-root semimetal by capping our void-channel network with large voids, thereby ensuring that the chiral symmetry of the squared Hamiltonian is not broken by the interfaces. We study the interfaces in a ribbon and a triangular metaparticle of the honeycomb-kagome lattice, and observe that topologically protected edge and corner states can be excited.

\section{Methodology}\label{sec:methodology}

\begin{figure}
    \centering
    \includegraphics{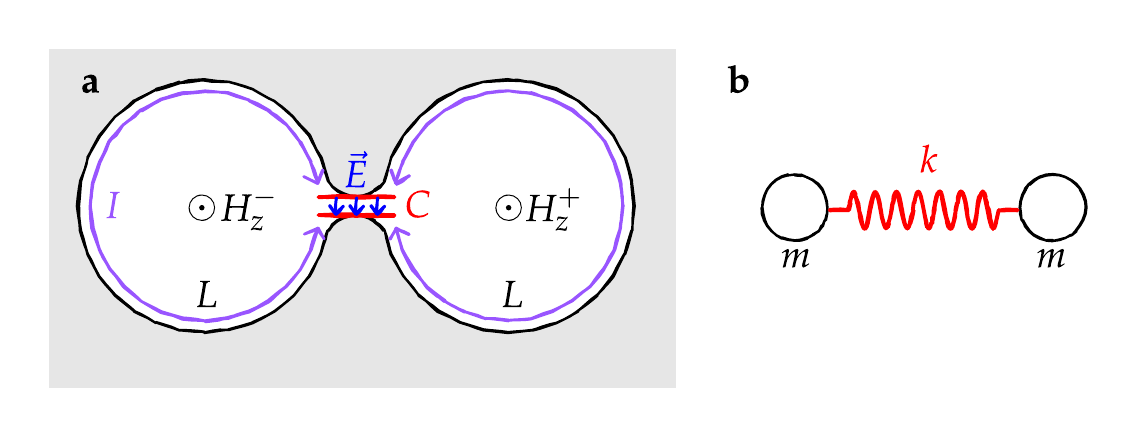}
    \caption{(a) Two voids of vacuum (white) and a narrow connecting channel embedded in perfect conductor (grey) will behave like a pair of inductors and a capacitor for transverse-electric polarised light at low frequencies and for narrow channels \cite{vanel2017asymptotic,vanel2018asymptotic}. The voids act as inductors with inductance $L$ because the currents, $I$, circulate around the surface of the voids and induce out-of-plane magnetic fields within the voids. We label the magnetic field within the left and right voids as $H_{z}^{-}$ and $H_{z}^{+}$, respectively. The channel acts as a capacitor because the difference of magnetic field across the channel generates an electric field, $\vec{E}$, across the gap. This analogy can be extended to larger networks of voids and channels. (b) The void-channel or inductor-capacitor network is also analogous to a mass-spring network, where masses $m$ are connected by spring constants $k$, and the masses oscillate in and out of the page.} \label{fig:void-channel-to-mass-spring}
\end{figure}

We apply the method of Vanel \emph{et al} \cite{vanel2017asymptotic} to map the TE-polarised Maxwell's equations, within networks of voids and channels formed between closely spaced perfect conductors (as shown in figure \ref{fig:void-channel-to-mass-spring}(a)), to equivalent networks of resonators in an asymptotically exact manner. The voids behave as inductors and the channels behave as capacitors; the difference in the out-of-plane magnetic field across a channel, $H_z^{+}-H_z^{-}$, results in an electric field, $\vec{E}$, perpendicular to the channel \cite{vanel2017asymptotic,vanel2018asymptotic}. Equivalently, we may consider a mass-spring network where the voids are mapped to masses and the channels are mapped to springs, as shown in figure \ref{fig:void-channel-to-mass-spring}(b).

The parameters of the discrete inductor-capacitor/mass-spring networks are not reliant upon lumped parameters and/or heuristic approximations; these approaches are common in electrical engineering as lumped circuit models \cite{sun2012}, or as optimisation with databases \cite{matlack_designing_2018}, and these successfully to take complex systems across to networks. The advantage of the alternative approach here is that the effective  
 parameters are simple and explicit. We proceed by utilising matched asymptotic expansions, Vanel \emph{et al} \cite{vanel2017asymptotic} demonstrated that the precise values of the masses and spring constants, corresponding to a particular network of voids and channels, can be determined in a remarkably simple manner at low frequencies and in the limit of narrow channels, $h/a\rightarrow0$. The masses are proportional to the area of the voids,
\begin{equation}\label{eq:masses}
    m_i = A_i \cdot m_0 / a^2,
\end{equation}
whilst the spring constants are a function of the half-width of the channel, $h$, in addition to the radius of curvature of the two sides of the channel, $R_1$ and $R_2$,
\begin{equation}\label{eq:spring-constant}
    k = \frac{1}{\pi} \sqrt{ \frac{h}{R_1} + \frac{h}{R_2} }.
\end{equation}

Equations \eqref{eq:masses} and \eqref{eq:spring-constant} allow us to accurately model the continuous void-channel network using the much simpler equations of motion of a discrete mass-spring system, without the need for any fine tuning or parameter fitting. The reverse mapping allows us to propose new photonic void-channel models where the coupling of the field between different voids is highly controllable. This afford us complete control over the symmetries of the Hamiltonian and, hence, suggests that void-channel networks could be a powerful platform for realising symmetry-protected photonic topological phases.

In this paper, the void-channel solutions were obtained using the open source finite element solver \texttt{FreeFem++} \cite{FreeFEM}; we used this to solve the Helmholtz equation for our void-channel model,
\begin{equation}
    \frac{\partial^2H_z}{\partial t^2} - c^2 \nabla^2H_z = 0,
\end{equation}
where $c$ is the speed of light in the space between the perfect conductor and where we notably applied Neumann boundary conditions along the surface of the perfect conductor \cite{vanel2017asymptotic,vanel2018asymptotic}. This was achieved using a modified version of a set of \texttt{FreeFem++} scripts that were originally used to model phononic crystals \cite{laude2015phononic}.

\section{Results}

\subsection{The Su-Schrieffer-Heeger chain}\label{sec:SSH}

\begin{figure}
    \centering
    \includegraphics{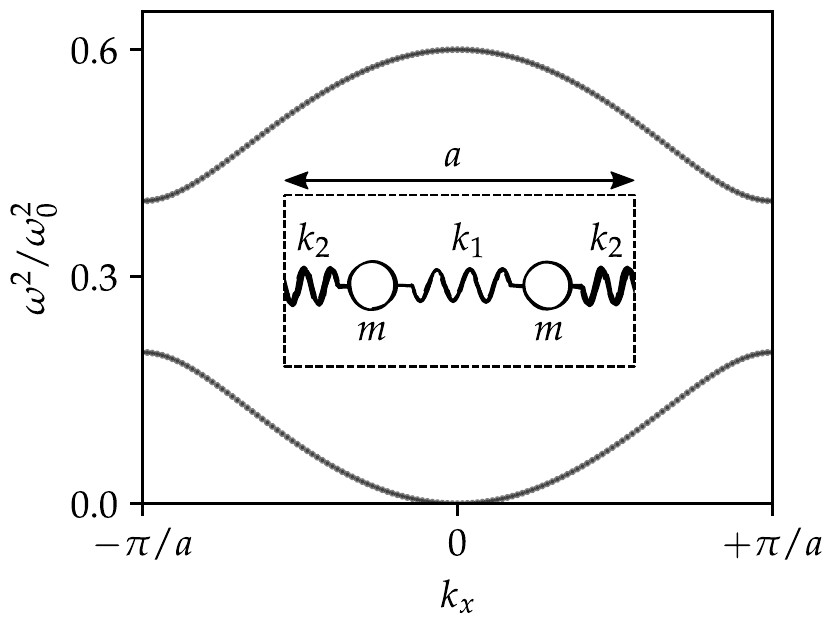}
    \caption{Squared frequency spectrum (normalised by $\omega_{0}=\sqrt{k_{0}/m_{0}}$) of a mass-spring chain with equal masses, $m=0.1m_{0}$, and alternating spring constants $k_{1}=0.01k_{0}$ and $k_{2}=0.02k_{0}$, as shown in the inset. The width of the unit cell is $a$. The spectrum closely resembles the energy spectrum of an electronic SSH tight-binding model \cite{su1980soliton,asboth2016short}, except that the squared frequency has been shifted by $(k_{1}+k_{2})/m=0.3\omega_{0}^{2}$.}
    \label{fig:SSH-mass-spring-bulk}
\end{figure}

To explore the symmetries of photonic void-channel networks, let us first consider the equations of motion of an SSH-like mass-spring network consisting of equal masses, $m$, and alternating spring constants, $k_1$ and $k_2$, as shown in the inset of figure \ref{fig:SSH-mass-spring-bulk},
\begin{equation}\label{eq:mass-spring-SSH-bulk}
    \left[\begin{array}{cc}
    \frac{k_{1}}{m}+\frac{k_{2}}{m} & -\frac{k_{1}}{m}-\frac{k_{2}}{m}e^{-ika}\\
    -\frac{k_{1}}{m}-\frac{k_{2}}{m}e^{+ika} & \frac{k_{1}}{m}+\frac{k_{2}}{m}
    \end{array}\right]\left[\begin{array}{c}
    \ket{u_{1}}\\
    \ket{u_{2}}
    \end{array}\right]=\omega^{2}(k)\left[\begin{array}{c}
    \ket{u_{1}}\\
    \ket{u_{2}}
    \end{array}\right].
\end{equation}
We see in figure \ref{fig:SSH-mass-spring-bulk} that the squared frequency spectrum of this chain resembles the energy spectrum of the electronic SSH tight-binding model. Unlike the tight-binding model the diagonal terms in equation \eqref{eq:mass-spring-SSH-bulk} are non-zero; however, as they are equal, they merely correspond to a simple shift of frequency by $\sqrt{k_1/m+k_2/m}$,
\begin{equation}
    \left[\begin{array}{cc}
    0 & -\frac{k_{1}}{m}-\frac{k_{2}}{m}e^{-ika}\\
    -\frac{k_{1}}{m}-\frac{k_{2}}{m}e^{+ika} & 0
    \end{array}\right]\left[\begin{array}{c}
    \ket{u_{1}}\\
    \ket{u_{2}}
    \end{array}\right]=\left(\omega^{2}(k)-\frac{k_{1}+k_{2}}{m}\right)\left[\begin{array}{c}
    \ket{u_{1}}\\
    \ket{u_{2}}
    \end{array}\right].
\end{equation}
This frequency shift allows for the chiral symmetry of the bulk equations to be preserved.

\begin{figure}
    \centering
    \includegraphics{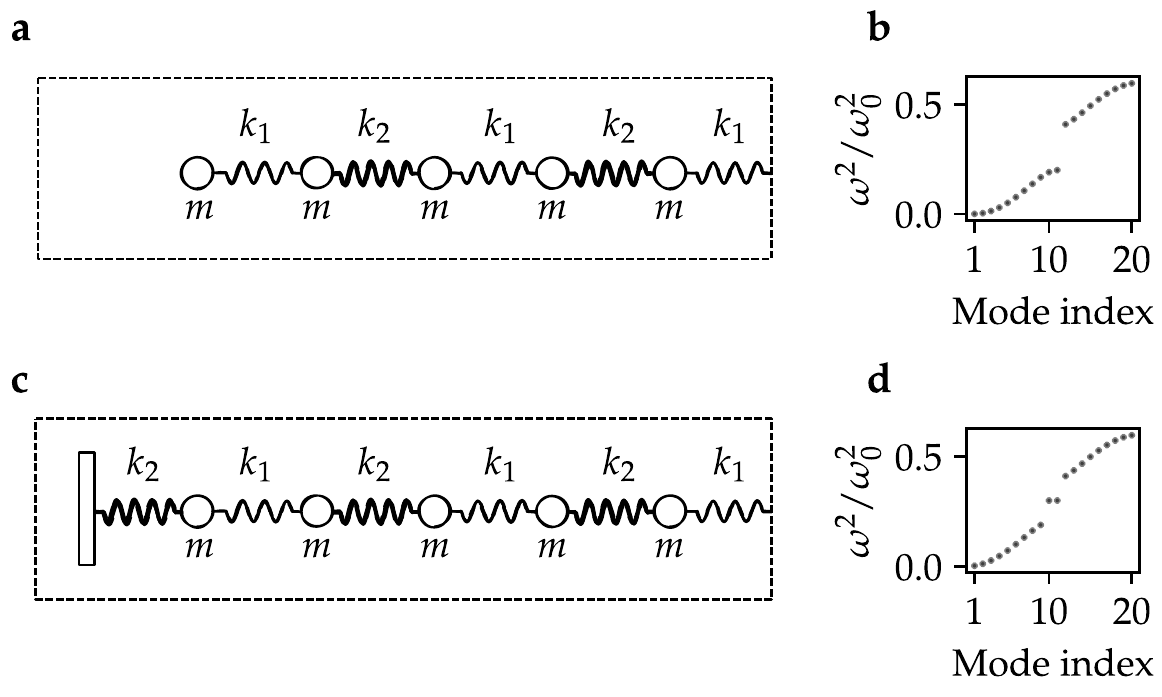}
    \caption{An SSH-like mass-spring chain with free and ``wall'' boundary conditions. The mass-spring chain consists of equal masses, $m$, connected by springs of alternating spring constants $k_{1}$ and $k_{2}$. (a‑b) Schematic and squared frequency spectrum of an SSH-like mass-spring chain with free boundary conditions. The chain is 20 masses long and we take $k_{1}=0.02k_0$, $k_{2}=0.01k_0$, and $m=0.1m_0$. Although the winding number of the bulk Hamiltonian is non-trivial, no edge states are observed because the free boundary condition breaks the chiral symmetry. (c‑d) Schematic and squared frequency spectrum of the same chain but with the ``wall'' boundary condition, where the edges of the SSH-like mass-spring chain are attached to immovable walls with springs of spring constant $k_{2}$. The wall boundary condition restores the chiral symmetry of the chain and symmetry-protected edge states are observed in the bulk band gap.}
    \label{fig:SSH-mass-spring-intro}
\end{figure}

The mass-spring model differs from the original SSH tight-binding model \cite{su1980soliton,asboth2016short} because the forces on the masses are proportional to the \emph{differences} of the mass displacements, whereas in tight-binding models the hopping is proportional to the wavefunction amplitudes themselves \cite{ni2019observation}. In particular, while the original SSH model is chiral symmetric for a finite chain with free boundary conditions, this is not the case for the mass-spring chain with free boundary conditions. The chiral symmetry is broken, in the latter, by the non-zero term along the diagonal of the matrix and therefore there are no topological edge states in the frequency spectrum, (see figure \ref{fig:SSH-mass-spring-intro}(b)),
\begin{equation}
    \left[\begin{array}{ccccc}
    -\frac{k_{2}}{m} & -\frac{k_{1}}{m}\\
    -\frac{k_{1}}{m} & 0 & -\frac{k_{2}}{m}\\
     & -\frac{k_{2}}{m} & 0 & -\frac{k_{1}}{m}\\
     &  & -\frac{k_{1}}{m} & 0 & \ddots\\
     &  &  & \ddots & \ddots
    \end{array}\right]\left[\begin{array}{c}
    u_{1}\\
    u_{2}\\
    u_{3}\\
    u_{4}\\
    \vdots
    \end{array}\right]=\left(\omega^{2}(k)-\frac{k_{1}+k_{2}}{m}\right)\left[\begin{array}{c}
    u_{1}\\
    u_{2}\\
    u_{3}\\
    u_{4}\\
    \vdots
    \end{array}\right].
\end{equation}
This distinction arises because the end masses are only connected to a solitary spring; we can restore the chiral symmetry by anchoring the chain to an immovable wall with a spring, of spring constant $k_2$, as shown in figure \ref{fig:SSH-mass-spring-intro}(c). The equations of motion of this chain with the ``wall'' boundary condition then become,
\begin{equation}
    \left[\begin{array}{ccccc}
    0 & -\frac{k_{1}}{m}\\
    -\frac{k_{1}}{m} & 0 & -\frac{k_{2}}{m}\\
     & -\frac{k_{2}}{m} & 0 & -\frac{k_{1}}{m}\\
     &  & -\frac{k_{1}}{m} & 0 & \ddots\\
     &  &  & \ddots & \ddots
    \end{array}\right]\left[\begin{array}{c}
    u_{1}\\
    u_{2}\\
    u_{3}\\
    u_{4}\\
    \vdots
    \end{array}\right]=\left(\omega^{2}(k)-\frac{k_{1}+k_{2}}{m}\right)\left[\begin{array}{c}
    u_{1}\\
    u_{2}\\
    u_{3}\\
    u_{4}\\
    \vdots
    \end{array}\right].
\end{equation}
We pictorially see, from figure \ref{fig:SSH-mass-spring-intro}(d), that the chiral symmetry is restored and, hence, SSH-like edge states emerge at the mid-gap frequency, $\sqrt{k_1/m+k_2/m}$.


\begin{figure}
    \centering
    \includegraphics{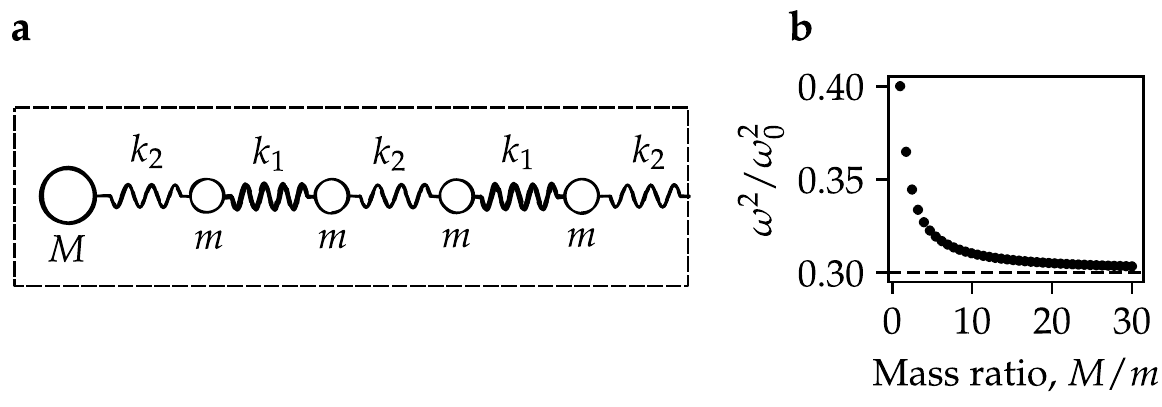}
    \caption{(a) The ``wall'' boundary condition of the SSH-like mass-spring chain, introduced in figure \ref{fig:SSH-mass-spring-intro}(b), can be approximated by replacing the wall with heavy masses, $M$. (b) The average frequency of the pair of edge states for the mass-spring chain capped with heavy masses (dots) compared with that for the mass-spring chain with the wall boundary condition (solid line), for the same values of $m$, $k_{1}$, and $k_{2}$ as before. Each chain contain 20 masses of mass $m$, and the capped chain has two capping masses of mass $M$ on each end for a total of 22 masses. For capping masses that are an order of magnitude heavier than the masses in the rest of the chain, $M \gtrsim 10m$, the squared frequencies of the edge states of the two chains agree within about 3\% error.}
    \label{fig:SSH-mass-spring-convergence-with-M}
\end{figure}

\begin{figure}
    \centering
    \includegraphics{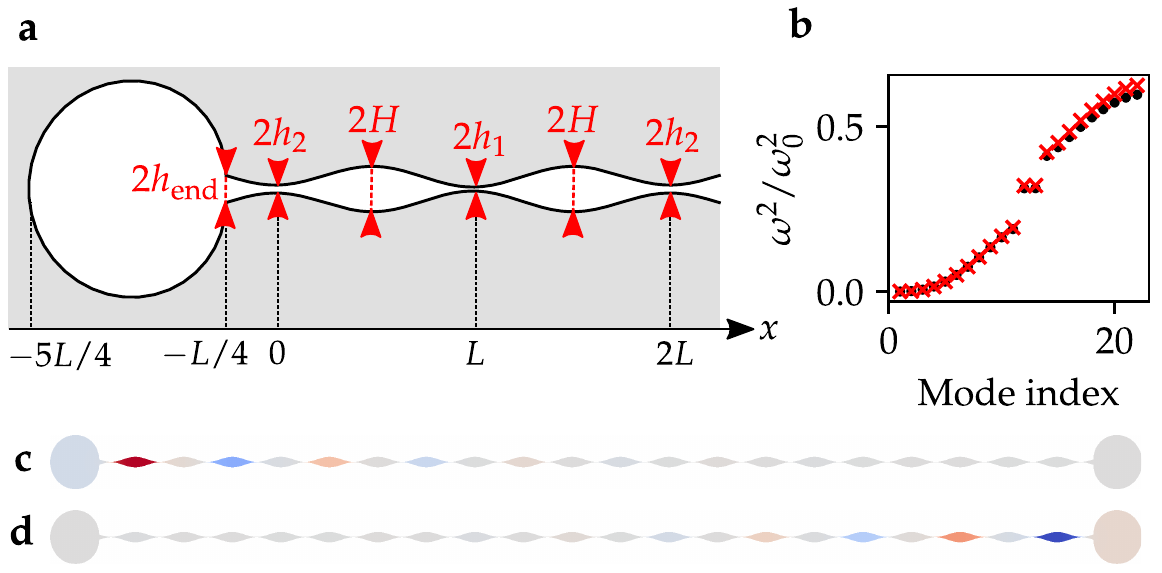}
    \caption{(a) Schematic of an SSH-like void-channel chain that is analagous to the capped mass-spring chain introduced in figure \ref{fig:SSH-mass-spring-convergence-with-M}. The grey region is perfect conductor, and the white is air. Only the beginning of the chain is shown. The bulk of the chain consists of equally sized voids of half-width $H$ and length $L$ connected by channels of alternating half-widths $h_{1}$ and $h_{2}$. The precise shapes of the voids and channels are defined in the main text. Note that the bulk region runs from $x=-L/4$ to $(N+1/4)L$ such that the curvature of the walls is well defined at the narrowest point of each channel. The half-width at the end of the bulk region is $h_{\mathrm{end}}$; the chain is then capped by larger voids that are roughly circular with diameter $L$. (b) Squared frequency spectrum of the void-channel chain (red crosses, $H=L/10$, $h_{1}=H/400$, $h_{2}=H/100$) and analagous mass-spring chain (black dots, $k_{1}=0.02$, $k_{2}=0.01$, $m=0.1m_0$, $M=\pi/4m_0$). The frequencies are normalised by $\omega_{0}=\sqrt{k_{0}/m_{0}}$ for the mass-spring model and $\omega_{0}=2\pi c_{0}/L$ for the void-channel model. The chains consist of 20 masses/voids (or 22 including the pair of larger masses/voids at the end). (c-d) The magnetic field (red positive, blue negative) of the two edge modes of the void-channel chain. We see that the chain preserves chiral symmetry well: the magnetic field is relatively weak in the large capping voids and each edge mode is well localised to just one sublattice.}
    \label{fig:void-channel-SSH}
\end{figure}

The wall boundary condition can be well approximated by capping mass-spring models with heavy masses, as shown in figure \ref{fig:SSH-mass-spring-convergence-with-M}. This allows us to propose a photonic analogue of the SSH model that consists of a one-dimensional network of voids and channels as shown in figure \ref{fig:void-channel-SSH}(a).

To emulate equal masses connected by alternating spring constants we require equally sized voids connected by relatively thin channels of alternating widths. A simple choice for the shape of the upper and lower walls of the geometry is
\begin{equation}
    y(x)=\pm\left[H\sin^{2}(\pi x/L)+h_{1}\sin^{2}(\pi x/2L)+h_{2}\cos^{2}(\pi x/2L)\right],
\end{equation}
where we take the half-width of the void as $H=L/10$, and the alternating half-widths of the channels as $h_1=H/400$ and $h_2=H/100$. Note that the upper and lower walls run from $x=-\tfrac{1}{4}L$ to $x=(N+\tfrac{1}{4})L$ to ensure that the radius of curvature of each channel from $x=0$ to $x=NL$ are well defined. The local radius of curvature of the walls of each channel is $R=L^2/(2\pi^2H)$. As $h_1,h_2 \ll H$, the area of each void in the bulk region is approximately $A_\mathrm{bulk} \approx 2\int_0^L H\sin^2(\pi x/L)\mathrm{d}x=HL$.

The walls are capped by roughly circular voids of diameter $L$. On the left hand side,
\begin{align}
    \label{eq:x_L}
    x_{\mathrm{L}}(\theta)	&=\frac{L}{2}\cos(\theta)-\frac{3}{4}L, \\
    y_{\mathrm{L}}(\theta) 	&=\frac{L}{2}\sin(\theta)+h_{\mathrm{end}}\cos(\theta/2),
\end{align}
for $\theta=[0,2\pi]$ and on the right hand side,
\begin{align}
    x_{\mathrm{R}}(\theta)	&=\frac{L}{2}\cos(\theta)+ \left(N+\frac{3}{4} \right)L,\\
    y_{\mathrm{R}}(\theta)	&=\frac{L}{2}\sin(\theta)-h_{\mathrm{end}}\sin(\theta/2),
    \label{eq:y_R}
\end{align}
for $\theta=[-\pi,\pi]$, where the $h_{\mathrm{end}}=y(-L/4)=y(NL+L/4)$ term is included to ensure that the walls of the geometry are continuous. As $h_{\mathrm{end}}\lesssim L$, we can take the area of the large capping voids as approximately $A_{\mathrm{cap}}=\pi L^{2}/4$. Note that this is a slight underestimate of the true area of the caps because we do not account for the region $-L/4\leq x\leq0$ or for the extra height of the void, described by Equations \eqref{eq:x_L}-\eqref{eq:y_R}, compared to a circle of diameter L.

The alternating spring constants of the corresponding mass-spring network are
\begin{align}
    k_{1}	&=\frac{1}{\pi}\sqrt{2h_{1}/R}\cdot k_{0}=0.01k_{0}, \\
    k_{2}	&=\frac{1}{\pi}\sqrt{2h_{2}/R}\cdot k_{0}=0.02k_{0},
\end{align}
the masses in the bulk of the chain are
\begin{align}
    m=A_{\mathrm{bulk}}\cdot m_{0}/L^{2}=0.1m_{0},
\end{align}
and the larger capping masses are
\begin{align}
    M=A_{\mathrm{cap}}\cdot m_{0}/L^{2}=\frac{\pi}{4}m_{0},
\end{align}
such that $M/m\approx8$, where $k_{1}$, $k_{2}$, $m$, and $M$ are as defined earlier in figure \ref{fig:SSH-mass-spring-convergence-with-M}. As the capping mass is roughly an order of magnitude larger than the bulk masses, we expect that the chiral symmetry is largely restored to the model. Indeed, figure \ref{fig:void-channel-SSH}(b) shows that the energy spectra of both the void-channel network (red) and the mass-spring network (black) behave as non-trivial SSH chains with a pair of topological edge states in the energy gap.

Overall, there is good agreement between the void-channel and mass-spring models; in particular we see similar band gaps and a pair of topological edge states in each model. The models agree well at lower frequencies, because the mapping between the models is valid for frequencies, below a cut-off frequency that scales as $\omega_{\mathrm{cutoff}}^{2}\sim1/h$ \cite{vanel2018asymptotic}. We shall explore this error, in greater detail, when we study the honeycomb-kagome lattice.

The edge states are largely localised on separate sublattices and decay quickly into the bulk, as shown for the void-channel model in figures \ref{fig:void-channel-SSH}(c‑d). It is intuitive that in the mass-spring model the heavy capping masses will oscillate with a smaller amplitude than the other masses, and we see that correspondingly the fields in the capping voids of the void-channel model are weak. The non-zero field within capping voids indicates that the chiral symmetry is not perfectly restored. This could be improved by increasing the size of the capping voids, however this does not seem necessary as the chiral symmetry violation is weak enough that the squared frequencies of the edge states remain well centered within the band gap.

\subsection{Flat edge states in the honeycomb-kagome lattice}\label{sec:honeycomb-kagome}

Having established that the chiral symmetry of the mass-spring/void-channel networks can be restored by capping the interfaces with sufficiently heavy masses/large voids, we now turn to a more complex case of a square-root semimetal where the topology is protected by the chiral symmetry of the honeycomb sector of the squared Hamiltonian \cite{arkinstall2017topological,mizoguchi2021square}. We shall see that despite the differences between the mass-spring/void-channel models and the original tight-binding model, capping the interfaces again allows the protecting symmetry to be restored and for the topological edge states to be observed.

\subsubsection{Tight-binding model}

\begin{figure}
    \centering
    \includegraphics{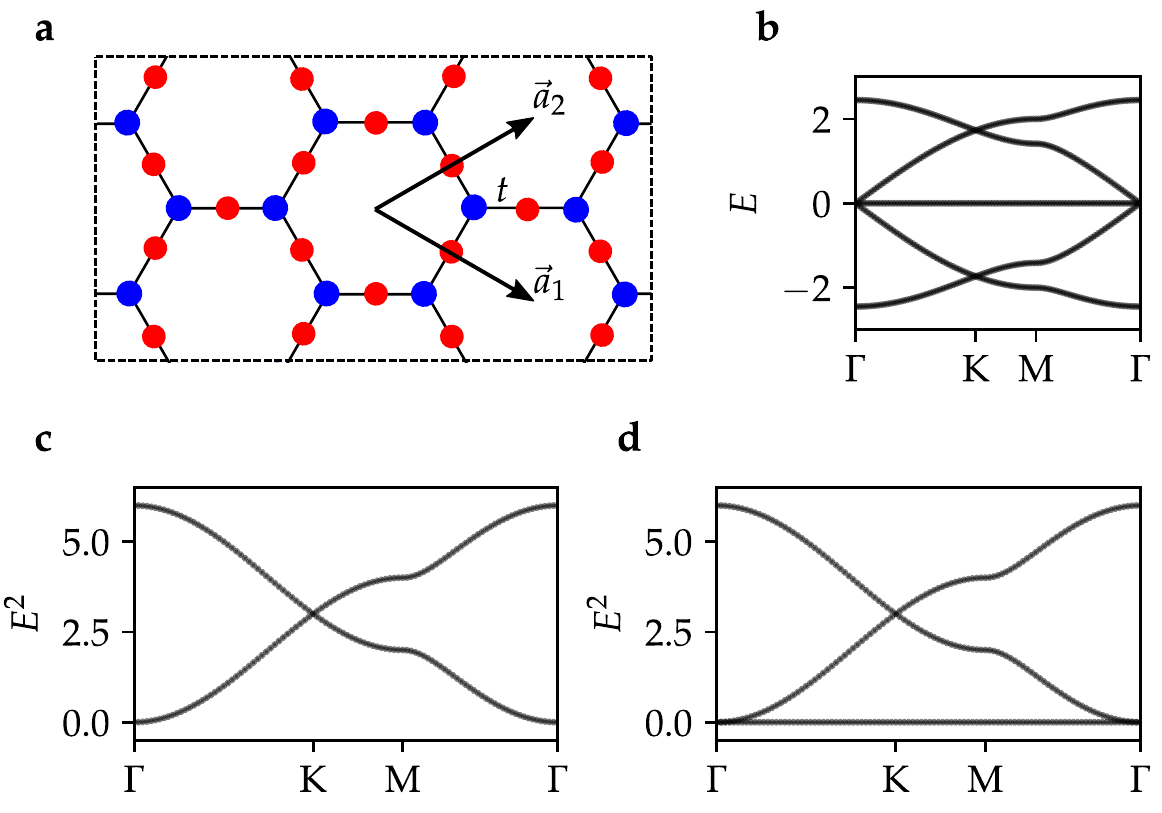}
    \caption{(a) Schematic of the tight-binding Hamiltonian of the honeycomb-kagome lattice, $\hat{H}^{\text{hk}}$, with nearest-neighbour hopping parameter $t$, as introduced by \cite{mizoguchi2021square}. The honeycomb and kagome sites are shown in blue and red, respectively, and $\vec{a}_{1}$ and $\vec{a}_{2}$ are the lattice parameters. (b) The energy spectrum of $\hat{H}^{\text{hk}}$ is symmetric about $E=0$ because of the chiral symmetry. The flat band and Dirac crossings at K are reminiscent of the nearest-neighbour tight-binding models on the honeycomb and kagome lattices, $\hat{H}^{\mathrm{h}}$ and $\hat{H}^{\mathrm{k}}$, respectively. This is because $\left(\hat{H}^{\text{hk}}\right)^{2}$ is block-diagonal, with the blocks proportional to the energy spectra of $\hat{H}^{\mathrm{h}}$ and $\hat{H}^{\mathrm{k}}$, up to a constant shift of energy, as shown in (c‑d) for the honeycomb and kagome sectors of $\left(\hat{H}^{\text{hk}}\right)^{2}$, respectively. The energy spectrum of $\hat{H}^{\text{hk}}$ therefore inherits features, including topology, from $\hat{H}^{\mathrm{h}}$ and $\hat{H}^{\mathrm{k}}$. In particular, $\hat{H}^{\mathrm{h}}$ is a topological semimetal, and $\hat{H}^{\text{hk}}$ is therefore known as a square-root topological semimetal \cite{arkinstall2017topological, mizoguchi2021square}.}
    \label{fig:honeycomb-kagome-TB}
\end{figure}

Before we introduce our mass-spring and void-channel models, let us review the tight-binding model introduced by Mizoguchi \emph{et al} \cite{mizoguchi2021square} and explain its topological origins. Figure \ref{fig:honeycomb-kagome-TB}(a) shows the nearest-neighbour tight-binding model of the honeycomb-kagome lattice, also known as the decorated honeycomb lattice. The Hamiltonian has a block off-diagonal form \cite{mizoguchi2021square},
\begin{equation}
    \label{eq:honeycomb-kagome-TB}
    \mat H_{\vec{k}}^{\text{hk}}\left[\begin{array}{c}
    u_{1}\\
    \vdots\\
    u_{5}
    \end{array}\right]=\left[\begin{array}{cc}
    \mat 0_{2\times2} & t\mat{\Psi}_{\vec{k}}^{\dagger}\\
    t\mat{\Psi}_{\vec{k}} & \mat 0_{3\times3}
    \end{array}\right]\left[\begin{array}{c}
    u_{1}\\
    \vdots\\
    u_{5}
    \end{array}\right],
\end{equation}
where $u_{1}$ and $u_{2}$ are amplitudes at the honeycomb sites, $u_{3}$, $u_{4}$, and $u_{5}$ are amplitudes at the kagome sites, $t$ is the hopping strength, $\mat 0_{n\times m}$ is an $n\times m$ matrix of zeros, and
\begin{equation}
    \mat{\Psi}_{\vec{k}}=\left[\begin{array}{cc}
    1 & 1\\
    e^{i\vec{k}\cdot\vec{a}_{1}} & 1\\
    e^{i\vec{k}\cdot\vec{a}_{2}} & 1
    \end{array}\right].
\end{equation}
As there are only hoppings between sites belonging to \emph{different} sublattices, this tight-binding model is chiral symmetric. The unitary chiral symmetry operator is
\begin{equation}
    \mat{\Gamma}=\left[\begin{array}{cc}
    \mat I_{2} & \mat 0_{2\times3}\\
    \mat 0_{3\times2} & -\mat I_{3}
    \end{array}\right],
\end{equation}
where $\mat I_{n}$ is an $n\times n$ identity matrix.

The energy bands of the tight-binding model, equation \eqref{eq:honeycomb-kagome-TB}, are shown in figure \ref{fig:honeycomb-kagome-TB}(b). The spectrum is symmetric about $E=0$ due to the chiral symmetry. Curiously, the spectrum contains features of both the underlying honeycomb and kagome lattices, such as the symmetry-protected Dirac cones at K \cite{schnyder2018lecture} and the flat band \cite{barreteau2017bird}. This is because equation \eqref{eq:honeycomb-kagome-TB} belongs to a class of Hamiltonians known as square-root Hamiltonians, meaning that the square of equation \eqref{eq:honeycomb-kagome-TB} is block diagonal,
\begin{equation}
    \label{eq:honeycomb-kagome-squared-TB}
    \left(\mat H_{\vec{k}}^{\text{hk}}\right)^{2}=\left[\begin{array}{cc}
    t^{2}\mat{\Psi}_{\vec{k}}^{\dagger}\mat{\Psi}_{\vec{k}} & \mat 0_{2\times3}\\
    \mat 0_{3\times2} & t^{2}\mat{\Psi}_{\vec{k}}\mat{\Psi}_{\vec{k}}^{\dagger}
    \end{array}\right]=\left[\begin{array}{cc}
     \mat H_{\vec{k}}^{\mathrm{h}}  &  \mat 0_{2\times3}\\
     \mat 0_{3\times2}  &  \mat H_{\vec{k}}^{\mathrm{k}} 
    \end{array}\right],
\end{equation}
where
\begin{equation}
    \label{eq:squared-honeycomb-sector}
    \mat H_{\vec{k}}^{\mathrm{h}}=\left[\begin{array}{cc}
    3t^{2} & (1+e^{+i\vec{k}\cdot\vec{a}_{1}}+e^{+i\vec{k}\cdot\vec{a}_{2}})t^{2}\\
    (1+e^{-i\vec{k}\cdot\vec{a}_{1}}+e^{-i\vec{k}\cdot\vec{a}_{2}})t^{2} & 3t^{2}
    \end{array}\right]
\end{equation}
is the tight-binding Hamiltonian of a honeycomb lattice with nearest-neighbour hopping strength $t^2$ and on-site potential $3t^2$, and
\begin{equation}
    \mat H_{\vec{k}}^{\mathrm{k}}=\left[\begin{array}{ccc}
    2t^{2} & (1+e^{-i\vec{k}\cdot\vec{a}_{1}})t^{2} & (1+e^{-i\vec{k}\cdot\vec{a}_{2}})t^{2}\\
    (1+e^{+i\vec{k}\cdot\vec{a}_{1}})t^{2} & 2t^{2} & (1+e^{-i\vec{k}(\vec{a}_{2}-\vec{a}_{1})})t^{2}\\
    (1+e^{+i\vec{k}\cdot\vec{a}_{2}})t^{2} & (1+e^{+i\vec{k}(\vec{a}_{2}-\vec{a}_{1})})t^{2} & 2t^{2}
    \end{array}\right]
\end{equation}
is the tight-binding Hamiltonian of a kagome lattice with nearest-neighbour hopping strength $t^{2}$ and on-site potential $2t^{2}$ \cite{mizoguchi2021square}. The squared energy spectrum of the honeycomb and kagome sectors, of equation \eqref{eq:honeycomb-kagome-squared-TB}, are shown in figures \ref{fig:honeycomb-kagome-TB}(c‑d), respectively.

Arkinstall \emph{et al} \cite{arkinstall2017topological} introduced a class of topological materials whose non-trivial topology is inherited from the squared Hamiltonian. They named these materials square-root topological insulators. The nearest-neighbour tight-binding model of the honeycomb lattice is a topological semi-metal. The honeycomb-kagome lattice is therefore a square-root topological semi-metal, with the non-trivial topology inherited from the honeycomb sector of the squared Hamiltonian \cite{mizoguchi2021square, delplace2011zak}. In the following sections, we introduce mass-spring and void-channel analogues of the square-root topological semimetal on the honeycomb lattice and study the symmetry protected edge states.

\subsubsection{Mass-spring and void-channel models}

\begin{figure}
    \centering
    \includegraphics{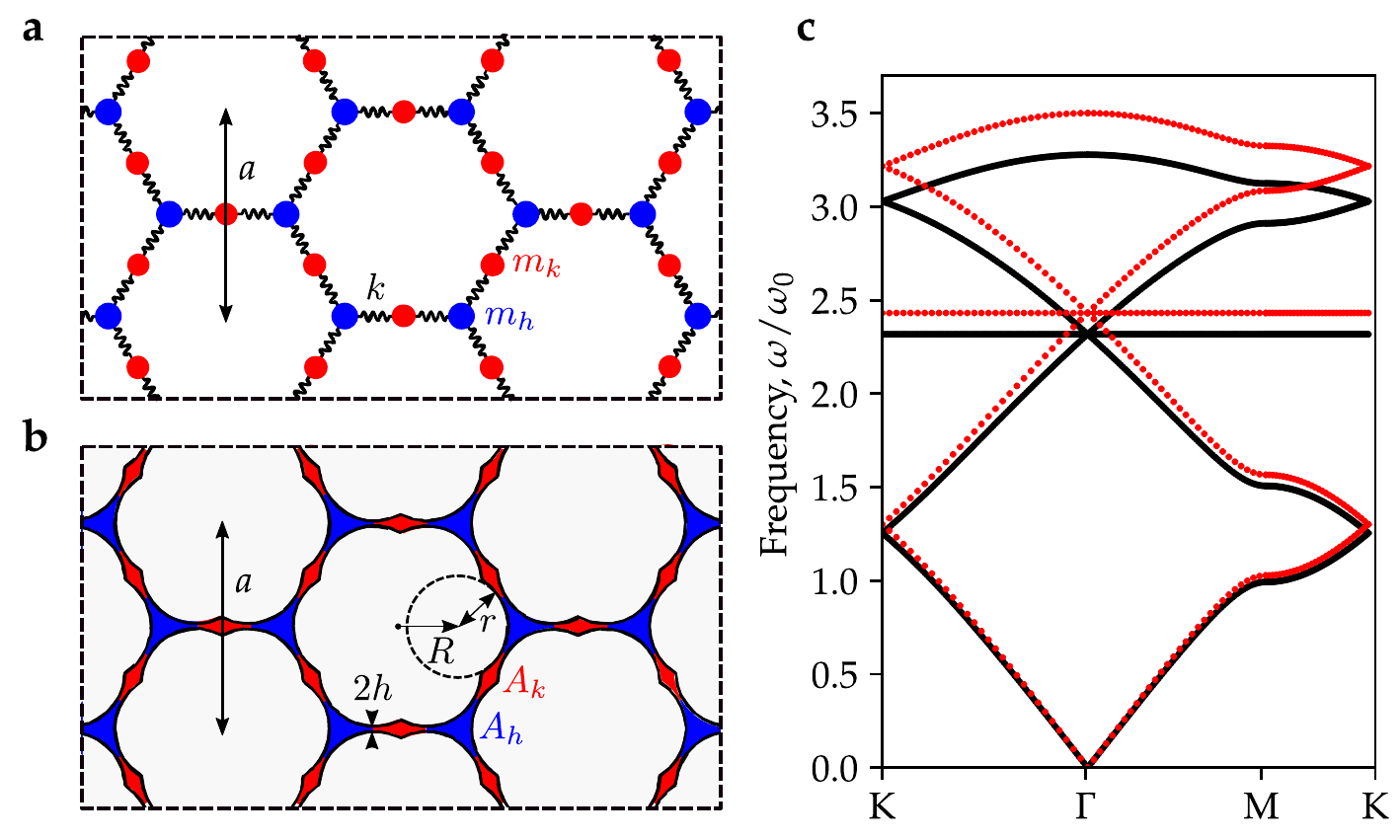}
    \caption{Mass-spring and void-channel models of a honeycomb-kagome lattice with nearest-neighbour coupling. (a) In the mass-spring model, the masses at honeycomb sites ($m_{h}$, blue) and kagome sites ($m_{k}$, red) are connected by springs of equal spring constants $k$. (b) In the void-channel model, voids and channels are formed between flower-shaped perfectly conducting particles arranged on a triangular lattice of lattice parameter $a$. The flower shapes consist of six cylinders of radius $r$ arranged in a ring of radius $R$. The voids at honeycomb the honeycomb sites have area $A_{h}$ (blue voids) and the voids at kagome sites have area $A_{h}$ (red voids). The channels have equal half-widths $h$. (c) Bulk frequency bands of the mass-spring model (black points, $m_{h}=0.01104m_{0}$, $m_{k}=0.00736m_{0}$, $k=0.01978k_{0}$) and the void-channel model (red points, $A_{h}=0.01104a^{2}$, $A_{k}=0.00736a^{2}$, $2h=0.001a$). The frequencies are normalised by $\omega_{0}=\sqrt{k_{0}/m_{0}}$ for the mass-spring model and $\omega_{0}=2\pi c_{0}/a$ for the void-channel model. There is good agreement between the two models, particularly at the lower frequencies. We chose the masses and areas such that $m_{h}/m_{k}=A_{h}/A_{h}=3/2$ in order that the models resemble the tight-binding model of figure \ref{fig:honeycomb-kagome-TB}(c) but with a shift of frequency, as discussed in the main text.}
    \label{fig:honeycomb-kagome-bulk}
\end{figure}

Figure \ref{fig:honeycomb-kagome-bulk}(a) shows a mass-spring model of the honeycomb-kagome lattice where the honeycomb masses, $m_h$, and kagome masses, $m_k$, are connected by springs of equal spring constant $k$,
\begin{equation}
    \label{eq:honeycomb-kagome-mass-spring-bulk}
    \left[\begin{array}{cc}
    \frac{3k}{m_{h}}\mat I_{2\times2} & -\frac{k}{m_{h}}\mat{\Psi}_{\vec{k}}^{\dagger}\\
    -\frac{k}{m_{k}}\mat{\Psi}_{\vec{k}} & \frac{2k}{m_{k}}\mat I_{3\times3}
    \end{array}\right]\left[\begin{array}{c}
    u_{1}\\
    u_{2}\\
    u_{3}\\
    u_{4}\\
    u_{5}
    \end{array}\right]=\omega^{2}(\vec{k})\left[\begin{array}{c}
    u_{1}\\
    u_{2}\\
    u_{3}\\
    u_{4}\\
    u_{5}
    \end{array}\right].
\end{equation}
First, note that the unsquared equations have eigenvalue $\omega^{2}$, and the squared equations would have eigenvalue $\omega^{4}$. Next, we note that in the mass-spring model the block-diagonal terms are non-zero and the two off-diagonal block matrices are scaled by different factors, namely $m_{h}$ and $m_{k}$. In a recent study of tight-binding and mass-spring honeycomb-kagome lattices, Mizoguchi \emph{et al} \cite{mizoguchi2021square} reproduced the tight-binding model by letting $m_{h}=m_{k}$ and setting the block-diagonal of the matrix to zero by adding a gravitional potential term in which the masses roll around in dents on a floor. Our interest is in mapping the mass-spring models to void-channel networks but no analogue of these dents for the void-channel network was apparent to us. Regardless, we shall demonstrate the squared tight-binding and mass-spring models are analagous. First, we decompose the unsquared mass-spring matrix equation as
\begin{equation}
    \left[\begin{array}{cc}
    \frac{3k}{m_{h}}\mat I_{2\times2} & -\frac{k}{m_{h}}\mat{\Psi}_{\vec{k}}^{\dagger}\\
    -\frac{k}{m_{k}}\mat{\Psi}_{\vec{k}} & \frac{2k}{m_{k}}\mat I_{3\times3}
    \end{array}\right]=\frac{\alpha k}{m_{0}}\mat I_{5\times5}+\left[\begin{array}{cc}
    +\frac{\beta k}{m_{0}}\mat I_{2\times2} & -\frac{k}{m_{h}}\mat{\Psi}_{\vec{k}}^{\dagger}\\
    -\frac{k}{m_{k}}\mat{\Psi}_{\vec{k}} & -\frac{\beta k}{m_{0}}\mat I_{3\times3}
    \end{array}\right],
\end{equation}
where
\begin{align}
    \alpha&=\frac{3/m_{h}+2/m_{k}}{2}m_{0},\\
    \beta&=\frac{3/m_{h}-2/m_{k}}{2}m_{0}.\label{eq:beta}
\end{align}
Taking the $\alpha k/m_{0}$ term to the right hand side of the equations of motion, we obtain
\begin{equation}
        \label{eq:freq-squared-minus-alpha}
        \left[\begin{array}{cc}
    +\frac{\beta k}{m_{0}}\mat I_{2\times2} & -\frac{k}{m_{h}}\mat{\Psi}_{\vec{k}}^{\dagger}\\
    -\frac{k}{m_{k}}\mat{\Psi}_{\vec{k}} & -\frac{\beta k}{m_{0}}\mat I_{3\times3}
    \end{array}\right]\left[\begin{array}{c}
    u_{1}\\
    \vdots\\
    u_{5}
    \end{array}\right]=\left(\omega^{2}(\vec{k})-\frac{\alpha k}{m_{0}} \right)\left[\begin{array}{c}
    u_{1}\\
    \vdots\\
    u_{5}
    \end{array}\right].
\end{equation}
Note that the equations of motion are only chiral symmetric about $\omega^{2}=\alpha k/m_{0}$ if we choose $2m_{h}=3m_{k}$ such that $\beta=0$.

The matrix of equation \eqref{eq:freq-squared-minus-alpha} squares to
\begin{align}
    \left[\begin{array}{cc}
    +\frac{\beta k}{m_{0}}\mat I_{2\times2} & -\frac{k}{m_{h}}\mat{\Psi}_{\vec{k}}^{\dagger}\\
    -\frac{k}{m_{k}}\mat{\Psi}_{\vec{k}} & -\frac{\beta k}{m_{0}}\mat I_{3\times3}
    \end{array}\right]^{2}&=\left(\frac{\beta k}{m_{0}}\right)^{2}\mat I+\left[\begin{array}{cc}
    \frac{k^{2}}{m_{k}m_{h}}\mat{\Psi}_{\vec{k}}^{\dagger}\mat{\Psi}_{\vec{k}} & \mat 0_{2\times3}\\
    \mat 0_{3\times2} & \frac{k^{2}}{m_{k}m_{h}}\mat{\Psi}_{\vec{k}}\mat{\Psi}_{\vec{k}}^{\dagger}
    \end{array}\right],
\end{align}
such that the ensuing squared equations of motion are
\begin{equation}
    \left[\begin{array}{cc}
    \frac{k^{2}}{m_{k}m_{h}}\mat{\Psi}_{\vec{k}}^{\dagger}\mat{\Psi}_{\vec{k}} & \mat 0_{2\times3}\\
    \mat 0_{3\times2} & \frac{k^{2}}{m_{k}m_{h}}\mat{\Psi}_{\vec{k}}\mat{\Psi}_{\vec{k}}^{\dagger}
    \end{array}\right]\left[\begin{array}{c}
    u_{1}\\
    \vdots\\
    u_{5}
    \end{array}\right]=\left[\left(\omega^{2}(\vec{k})-\frac{\alpha k}{m_{0}}\right)^{2}-\left(\frac{\beta k}{m_{0}}\right)^{2}\right]\left[\begin{array}{c}
    u_{1}\\
    \vdots\\
    u_{5}
    \end{array}\right],
\end{equation}
which is analagous to the squared tight-binding equation \eqref{eq:honeycomb-kagome-squared-TB} with $t^{2}\leftrightarrow k^{2}/(m_{k}m_{h})$ and $E^{2}\leftrightarrow\left(\omega^{2}(\vec{k})-\frac{\alpha k}{m_{0}}\right)^{2}-\left(\frac{\beta k}{m_{0}}\right)^{2}$. Note that for the squared equations of motion, $\beta\neq0$ simply corresponds to another shift in frequency and does not break any symmetries of the squared equations.

Next, we propose a photonic analogue of this mass-spring network. Our network of voids and channels is formed between ``flower'' shaped particles of perfect conductors arranged on a triangular lattice with lattice parameter $a$; each flower consists of six cylinders of radius $r$ that are distributed along a ring of radius $R$ (see figure \ref{fig:honeycomb-kagome-bulk}(b)). The voids at the honeycomb and kagome sites (shown in blue and red, respectively) are connected by narrow channels of half-width $h$, where
\begin{equation}
    R\cos\frac{\pi}{6}=\frac{a}{2}-r-h.
\end{equation}
We fixed the surface-to-surface gap as 2h=a/1000, this is similar to the ratio we used for the SSH model in the previous section. While this is quite a small ratio of $h/a$, we see in \ref{app:gap-convergence} that less extreme ratios of $h/a$ would also be viable. For a given value of $r$, we can then determine $R=(a/2-h-r)/\cos(\pi/6)$ and numerically calculate the areas of the honeycomb and kagome voids, $A_{h}$ and $A_{k}$. We settled on $r=0.259a$, for which $R=0.27771a$, $A_{h}=0.01104a^{2}$, and $A_{k}=0.00736a^{2}$, such that $A_{h}/A_{k}=3/2$. As shown in figure \ref{fig:honeycomb-kagome-bulk}(c), there is good agreement between the system of voids and channels (red) and the discrete system of masses and springs (black) where
\begin{align}
    m_{h}&=A_{h}\cdot m_{0}/a^{2}=0.01104m_{0},\\m_{k}&=A_{k}\cdot m_{0}/a^{2}=0.00736m_{0},\\k&=\frac{1}{\pi}\sqrt{\frac{2h}{r}}=0.01978k_{0}.
\end{align}
We also verify in \ref{app:gap-convergence} that the agreement between the mass-spring and void-channel networks improves as the channels are made more narrow; this is in line with expectations from the asymptotic model \cite{vanel2017asymptotic}. We have chosen these particular parameters such that $A_{h}/A_{k}=m_{h}/m_{k}=3/2$ and therefore $\beta=0$ in order that the mass-spring and void-channel models more closely resemble the tight-binding model shown in figure \ref{fig:honeycomb-kagome-TB}. When we study the edge states in the next section, we shall see that the topological edge states persist even if $m_{h}/m_{k}\neq3$ and $\beta\neq0$.

Now that we have introduced our photonic geometry, let us compare and contrast our work with some recent studies of the honeycomb-kagome lattice in photonics and acoustics. Maimaiti \emph{et al} \cite{maimaiti2020microwave} studied the response of a triangular array of metallic cylinders to microwave radiation. They noted that the voids between the cylinders lie on a honeycomb-kagome lattice and they used Monte Carlo methods to fit their model to a honeycomb-kagome tight-binding model. Crucially, however, the cylinders were not closely spaced and the authors did not consider any topological aspects of the array; it is likely that the quality of the edge states in this system would be reduced by longer-ranged coupling between voids and their next-nearest-neighbours. On the other hand, Yan \emph{et al} \cite{yan2020acoustic} studied a honeycomb-kagome array of acoustic resonators connected by narrow channels and considered the symmetry protected topology. However, in their work the width of the channels were alternated to produce a square-root topological insulator where the topology was inherited from the breathing kagome sector of the squared Hamiltonian, whereas we study the lattice with equal channel widths, which is akin to the mass-spring/tight-binding models of Mizoguchi \emph{et al} \cite{mizoguchi2021square}, where the non-trivial topology is inherited from the honeycomb sector of the squared Hamiltonian.

\subsubsection{Edge states in a ribbon}

\begin{figure}
    \centering
    \includegraphics{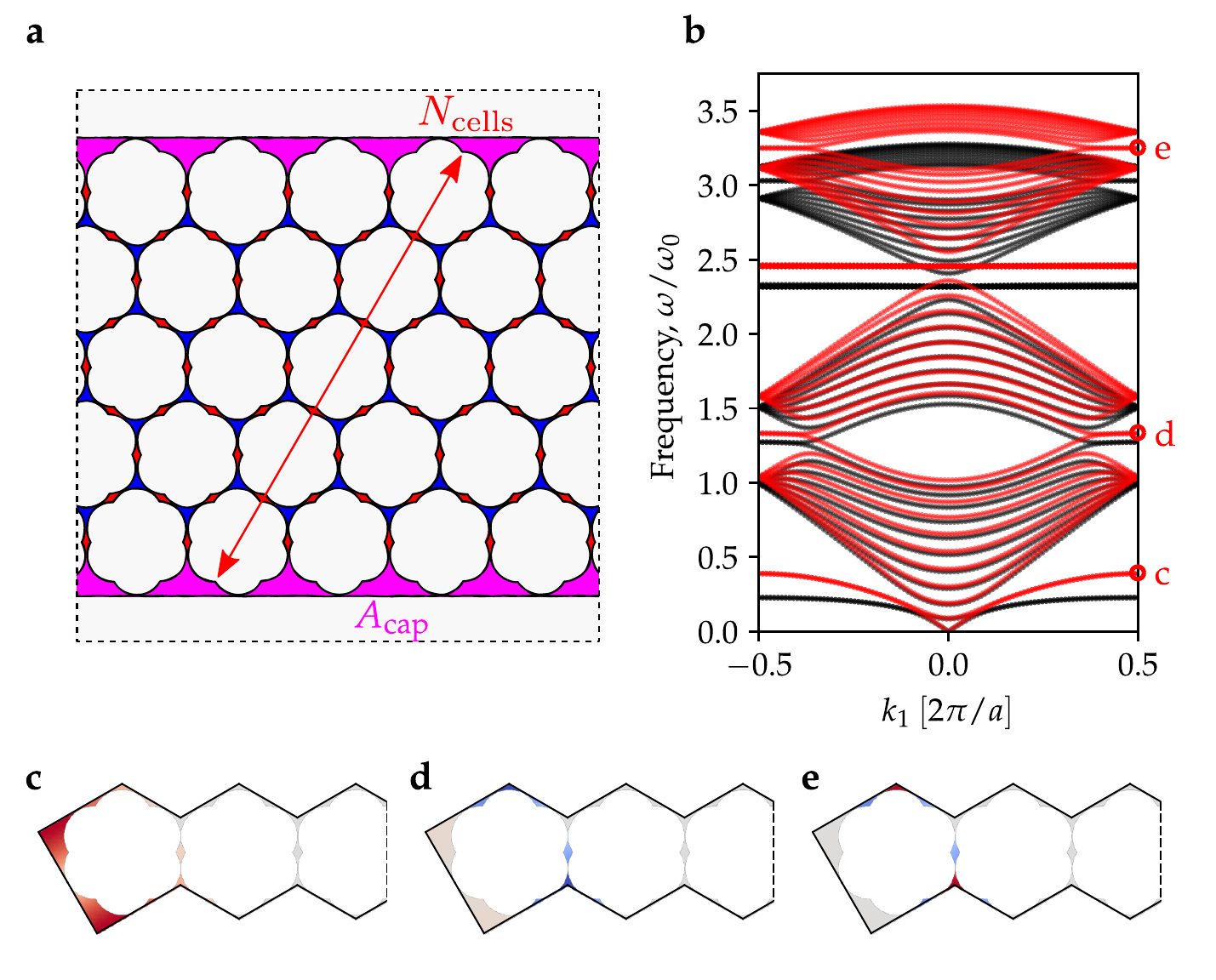}
    \caption{(a) Schematic of a ribbon of the honeycomb-kagome void-channel network introduced in figure \ref{fig:honeycomb-kagome-bulk}, but terminated by slabs of perfect conductor at the top and bottom. The large magenta voids at the boundary have area $A_{\mathrm{cap}}$ and reduce the chiral symmetry breaking at the interfaces. (b) Frequency bands of a ribbon that is $N_{\mathrm{cells}}=10$ unit cells long. The frequencies are normalised by $\omega_{0}=\sqrt{k_{0}/m_{0}}$ for the mass-spring model and $\omega_{0}=2\pi c_{0}/a$ for the void-channel model. (c-e). Visualisations of the labelled eigenmodes in panel b. For each shown here, this is also an energy degenerate inversion symmetric partner at the other edge. (c) The lowest pair of bands are excitations in the large voids, whereas (d-e) are topological edge states protected by the chiral symmetry of the squared Hamiltonian.}
    \label{fig:honeycomb-kagome-ribbon}
\end{figure}

In order to produce topological edge states, we must introduce interfaces in a manner that preserves (i) the block-diagonal nature of the squared equations and (ii) the chiral symmetry of the honeycomb sector of the squared equations \cite{mizoguchi2021square}. When we take the square of the tight-binding model with free boundary conditions, the sites at the edge of the model gain a different onsite potential when compared to the sites of the same sublattice, all be it, in the bulk. In order to retain the chiral symmetry of the honeycomb lattice, we impose that the kagome sites, in the tight-binding model, are located at the edge of the model. The edges of the mass-spring model therefore consist of kagome sites capped by heavy masses to emulate the free boundary condition.

Figure \ref{fig:honeycomb-kagome-ribbon}(b) shows a comparison between the void-channel model (red) and the mass-spring model (black, with the same values of $k$, $m_h$, and $m_k$ as before, and capped with masses of $M=A \cdot m_0/a^2=0.12247m_0$ at the honeycomb sites along the interface). As before, there is good agreement between the mass-spring model and void-channel models however the accuracy decreases as the frequency increases. We observe several new edge states, that were not present in the bulk eigenmodes; these are marked, in the dispersion curves, by the red circles c, d and e in figure \ref{fig:honeycomb-kagome-ribbon}(b), and visualised in figures \ref{fig:honeycomb-kagome-ribbon}(c-e), respectively.

We see from figure \ref{fig:honeycomb-kagome-ribbon}(c) that the lowest pair of bands correspond to excitations within the large voids. As we increase the size of the capping voids, these bands would flatten to zero frequency. On the other hand, figures \ref{fig:honeycomb-kagome-ribbon}(d-e) show the pair of topological edge states arising from the non-trivial topology of the honeycomb sector of the squared equations \cite{mizoguchi2021square}.

If the squared system was exactly chiral symmetric then the topological edge states should be flat \cite{mizoguchi2021square}; instead there is a slight tilt indicating a weak symmetry breaking. Interestingly, the field within the large voids is weaker for the higher energy eigenmode in figure \ref{fig:honeycomb-kagome-ribbon}(e), suggesting that the chiral symmetry of the squared system is better preserved at the higher frequencies. This is because the frequency and character of the unwanted excitation in the capping voids (see figure \ref{fig:honeycomb-kagome-ribbon}(c)) is more similar in character to the edge state with lower frequency (figure \ref{fig:honeycomb-kagome-ribbon}(d), where honeycomb and kagome sites are in phase) than the edge state with higher frequency (figure \ref{fig:honeycomb-kagome-ribbon}(e), where honeycomb and kagome sites are out of phase). The unwanted mode therefore hybridises more strongly with the lower frequency edge state.

Although we have chosen $m_h/m_k=3$, such that the mass-spring and void-channel models more closely resemble the tight-binding model of figure \ref{fig:honeycomb-kagome-TB}(e), we verify in \ref{app:chiral-sym-of-squared} that the edge states persist even if $m_{h}/m_{k}\neq 3$ such that $\beta\neq0$. We also verify in \ref{app:free-vs-wall} that the edge states are not protected without the presence of the large capping masses/voids which restore the chiral symmetry of the squared equations of motion.

\subsubsection{Edge and corner states in a triangular metaparticle}

\begin{figure}
    \centering
    \includegraphics{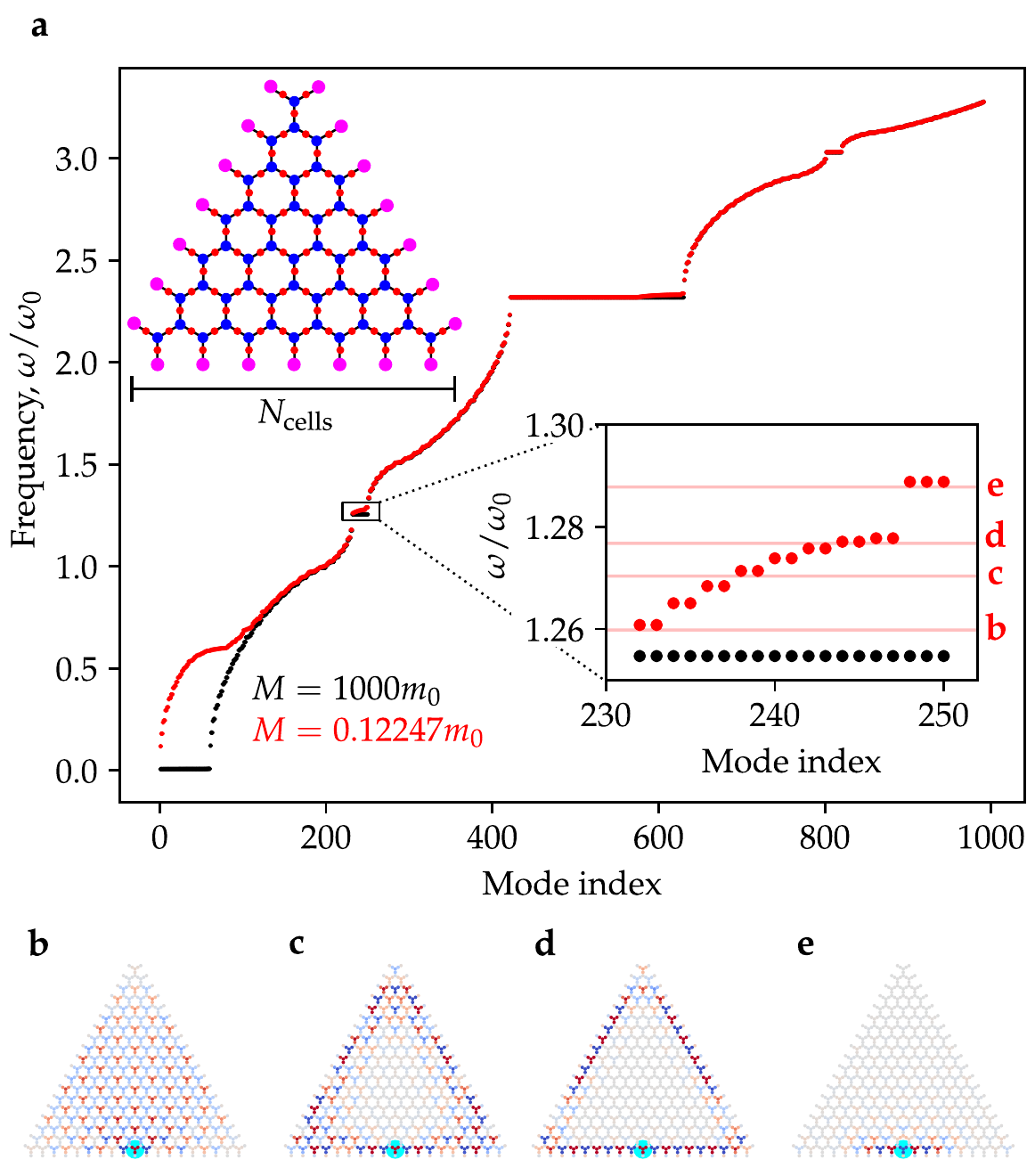}
    \caption{(a) Frequency eigenspectrum of a triangular metaparticle built from the mass-spring honeycomb-kagome network introduced in figure \ref{fig:honeycomb-kagome-bulk}. The edges are capped with extremely heavy masses ($M=1000m_{0}$, black points) or realistic masses ($M=0.12247m_{0}$, as in figure \ref{fig:honeycomb-kagome-ribbon}). The frequencies are normalised by $\omega_{0}=\sqrt{k_{0}/m_{0}}$. The upper left inset shows a schematic of the triangular metaparticle with the heavy masses shown in magenta. The schematic shows a particle with $N_{\mathrm{cells}}=7$ unit cells along each edge, but $N_{\mathrm{cells}}=19$ was used in the calculations. The lower right inset shows a zoom of the lower frequency set of edge and corner states. (b)‑(e) show steady state fields, upon driving the system with a harmonic force at the honeycomb site at the center of the highlighted blue region, at the frequencies indicated in the lower right inset of panel a.}
    \label{fig:honeycomb-kagome-triangle}
\end{figure}

We now study corner and edge states in a large but finite ``triangular metaparticle'' of the honeycomb-kagome lattice, as shown in the upper-left inset of figure \ref{fig:honeycomb-kagome-triangle}(a). Having established the validity of the mass-spring model for the bulk and at the edges, we model the system using only the discrete mass-spring equations as these are more accessible, far faster to solve and still retain the crucial physics we are interested in. As with the ribbon, we cap the ends with heavy masses at honeycomb sites to reduce the breaking of the chiral symmetry in the honeycomb sector of the squared equations.

The main panel of figure \ref{fig:honeycomb-kagome-triangle}(a) shows the energy spectrum of a triangular metaparticle with $N_\mathrm{cells}=19$ unit cells along each edge for a realistic capping mass ($M=0.12247m_0$, red) and for a very large capping mass where the chiral symmetry of the honeycomb sector of the squared equations is near-perfectly restored ($M=1000m_0$, black). We identify the large flat region of eigenmodes at $\omega\approx2.4\omega_0$ as the bulk flat band inherited from the kagome lattice, and the smaller flat regions of eigenmodes at $\omega\approx1.25\omega_0$ and $\omega\approx3.1\omega_0$ as the topological edge states inherited from the honeycomb lattice. The lower-right inset of figure \ref{fig:honeycomb-kagome-triangle}(a) shows the energy eigenmodes of the lower frequency edge state in more detail. Although the edge state is extremely flat, for the unrealistically large value of $M$, there is an advantage to using a more realistic value of $M$ for which the protecting symmetry is weakly broken.

\begin{figure}
    \centering
    \includegraphics{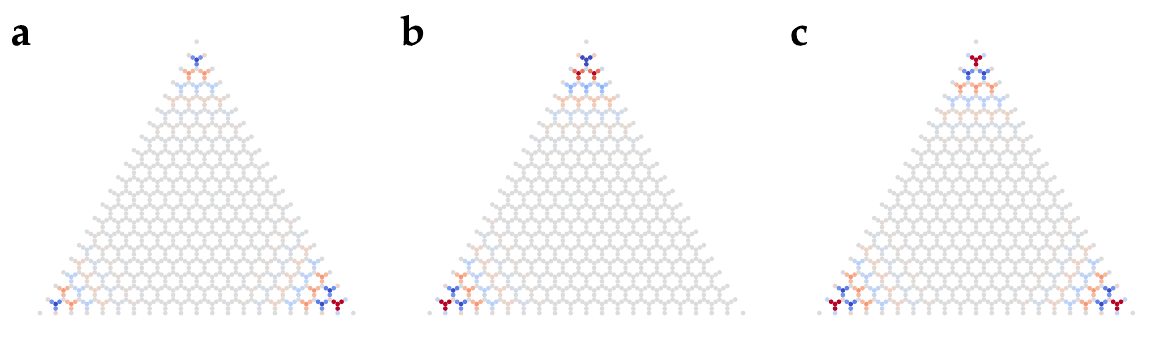}
    \caption{Visualisation of the triply degenerate eigenmodes of the triangular metaparticle for the mode indices (a) 248, (b) 249, (c) 250 of figure \ref{fig:honeycomb-kagome-triangle}(a). The fields reveal that these eigenmodes are corner states. The modes are degenerate because of the $\mathrm{C}_{3}$ symmetry of the triangle; the eigensolver has therefore returned arbitrary linear superpositions of the three corner eigenmodes.}
    \label{fig:corner-states}
\end{figure}

Figures \ref{fig:honeycomb-kagome-triangle}(b-e) show the steady-state solutions of the triangular mass-spring metaparticles capped by realistic masses and driven by time-harmonic forces, centered at the honeycomb sites highlighted in light blue, for the four frequencies labelled in the lower-right inset of figure \ref{fig:honeycomb-kagome-triangle}(a). Note that we forced the system at frequencies just below the resonances because the energy of the closed mass-spring system solution diverges if we drive exactly at a resonant frequency. In figure \ref{fig:honeycomb-kagome-triangle}(b) the energy propagates freely through the particle. This is because the two eigenmodes that are closest to the driving frequency are actually bulk eigenmodes corresponding to the Dirac cones at $\mathrm{K}$ and $-\mathrm{K}$ of figure \ref{fig:honeycomb-kagome-bulk}(c). In figure \ref{fig:honeycomb-kagome-triangle}(c-d) we see that the energy propagates around the edge of the particle but not into the bulk. As the energy increases, the modes become more localised to the edges. Figure \ref{fig:honeycomb-kagome-triangle}(e) shows that the field is localised in all directions, when driving at the edge of the triangle, at the frequency residing slightly below the group of triply degenerate eigenstates. This is because these eigenstates are corner eigenstates, as shown in figure \ref{fig:corner-states}. Crucially, the weak breaking of the chiral symmetry of the squared equations has lifted the degeneracies between the bulk states and the edge/corner states, allowing these to be excited at different frequencies.

\section{Conclusions}

In this paper we have shown that networks of voids and narrow connecting channels between perfect conductors are a promising platform for mimicking chiral or square-root topological tight-binding models within photonics. This was done by mapping the tight-binding models to mass-spring models, and then mapping these mass-spring models to their asymptotically exact continuum analogue \cite{vanel2017asymptotic,vanel2018asymptotic}, comprised of void-channel networks.

We found that although introducing interfaces to the mass-spring/void-channel networks could break the symmetries that protected the topological edge states, these symmetries could be restored by capping the interfaces with heavy masses/large voids. We were able to create a photonic analogue of the 1D SSH model \cite{asboth2016short} with a chain of equally sized voids connected by channels of alternating widths, and a photonic analogue of a square-root topological semimetal \cite{arkinstall2017topological,mizoguchi2021square} with voids positioned on a honeycomb-kagome lattice and narrow channels connecting the nearest-neighbour voids.

More broadly, we hope that the asymptotic network approximations espoused here will provide a direct mapping to other complex photonic crystal phenomena, including and beyond topological physics. Discrete models are able to encompass highly non-trivial phenomenology and hence our approach provides a systematic and simplified route to engineer exotic responses in continuum photonic structures in an asymptotically exact manner.




\clearpage

\section{Acknowledgements}

S.J.P.\ acknowledges his studentship from the Centre for Doctoral Training on Theory and Simulation of Materials at Imperial College London funded by EPSRC Grant Number EP/L015579/1.  The support of the UK EPSRC through grants EP/L024926/1 is acknowledged by RVC and MM  as is that of the ERC H2020 FETOpen project BOHEME  under grant agreement No. 863179. 

\appendix

\section{Validity of the mass-spring networks as an analogue of the void-channel networks}\label{app:gap-convergence}

\subsection{Convergence with decreasing gap size}

\begin{figure}
    \centering
    \includegraphics{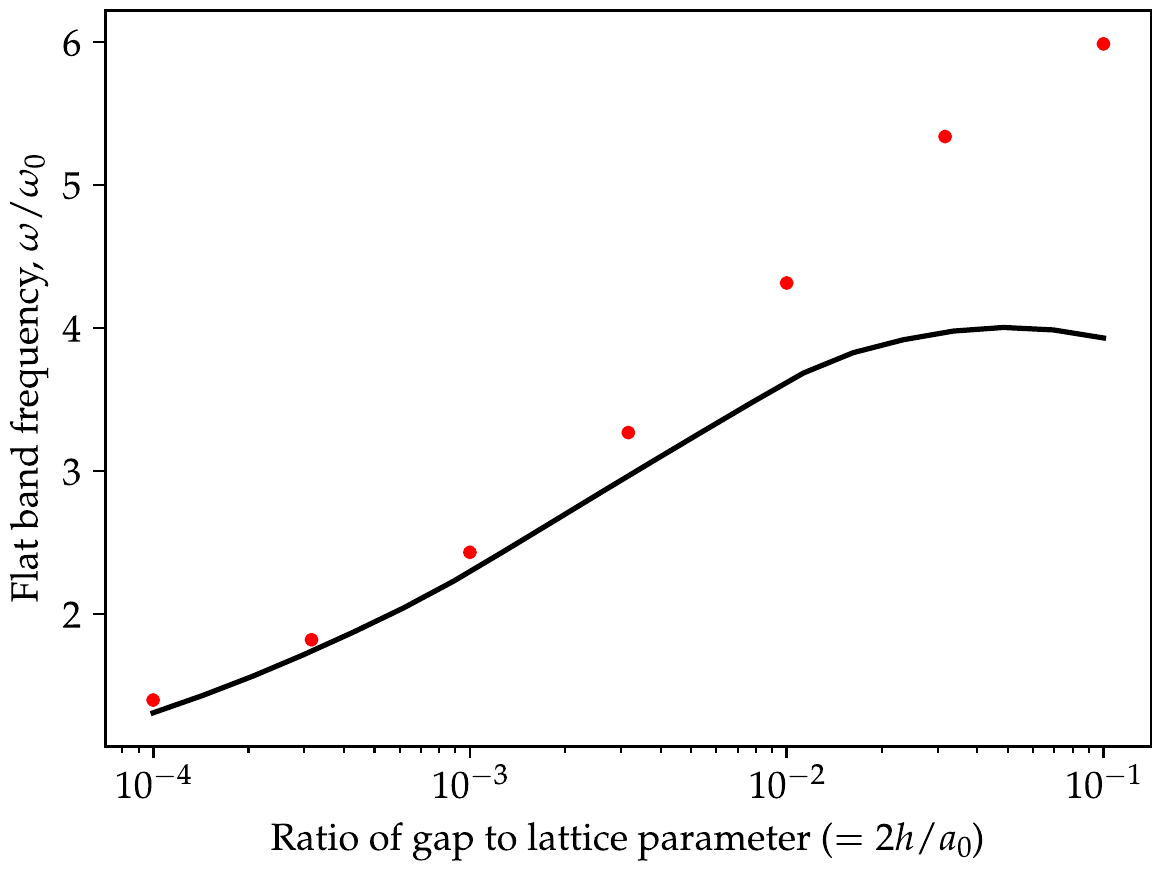}
    \caption{Normalised frequency of the flat band of honeycomb-kagome networks of masses and springs (black line) and the equivalent networks of voids and channels (red points). The agreement between the two models increases as the gap size is decreased.}
    \label{fig:gap-convergence}
\end{figure}

Figure \ref{fig:gap-convergence} shows the frequency of the flat band as a function of gap size in the honeycomb-kagome network of masses and springs (black line) and voids and channels (red points) that were originally introduced in figure \ref{fig:honeycomb-kagome-bulk}. The agreement between the two models increases as the gap size decreases, as expected from the asymptotic analysis of Vanel \emph{et al} \cite{vanel2017asymptotic,vanel2018asymptotic}.

\subsection{Operating frequencies and length scales}

Let us consider the feasibility of manufacturing the honeycomb-kagome network of voids and channels, and the frequencies and length scales at which it could operate. We must find a balance between the size of the gaps, the size of the particles, the frequencies at which the edge states occur, and the frequencies at which the metals may be treated as perfect conductors.

For example, let us consider the parameters required to obtain edge states at $\omega=\SI{1}{\tera\hertz}$. The lower frequency edge states have a normalised frequency of $\omega / \omega_0 \approx 1.3$, where $\omega_0=2\pi c_0/a$ and $c_0$ is the speed of light in vacuum, corresponding to a lattice parameter of $a\approx1.3\cdot 2\pi c_0/\omega=\SI{2.4}{\milli\meter}$. This would correspond to channel half-widths of $h=a/2000=\SI{1.2}{\micro\meter}$ using the ratio from earlier, although we have seen in the previous section that this could be relaxed somewhat without the mapping between the void-channel and mass-spring networks breaking down. Both $a$ and $h$ are orders of magnitude greater than the skin depth of gold which is on the order of $\SI{50}{\nano\meter}$ for $\omega=\SI{1}{\tera\hertz}$, and it would therefore be reasonable to treat the gold particles as perfectly conducting.

\section{No edge states in mass-spring/void-channel networks with free boundary conditions}\label{app:free-vs-wall}
\begin{figure}
    \centering
    \includegraphics{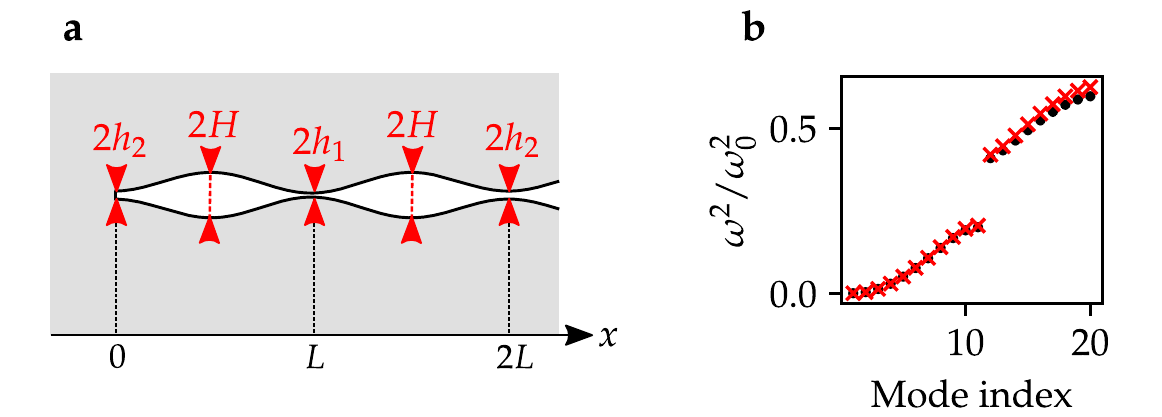}
    \caption{(a) Schematic of the same void-channel chain as in figure \ref{fig:void-channel-SSH} but without the large capping voids. (b) There is good agreement between the squared frequency spectrum of the void-channel chain (red crosses) and the corresponding mass-spring chain with free boundary conditions (black points). The frequencies are normalised by $\omega_{0}=\sqrt{k_{0}/m_{0}}$ for the mass-spring model and $\omega_{0}=2\pi c_{0}/L$ for the void-channel model. Without the large capping voids/heavy capping masses chiral symmetry is broken at the edges of these chains and there are no topological edge states in the band gap.}
    \label{fig:SSH-void-channel-broken-chiral-symmetry}
\end{figure}

\begin{figure}
    \centering
    \includegraphics{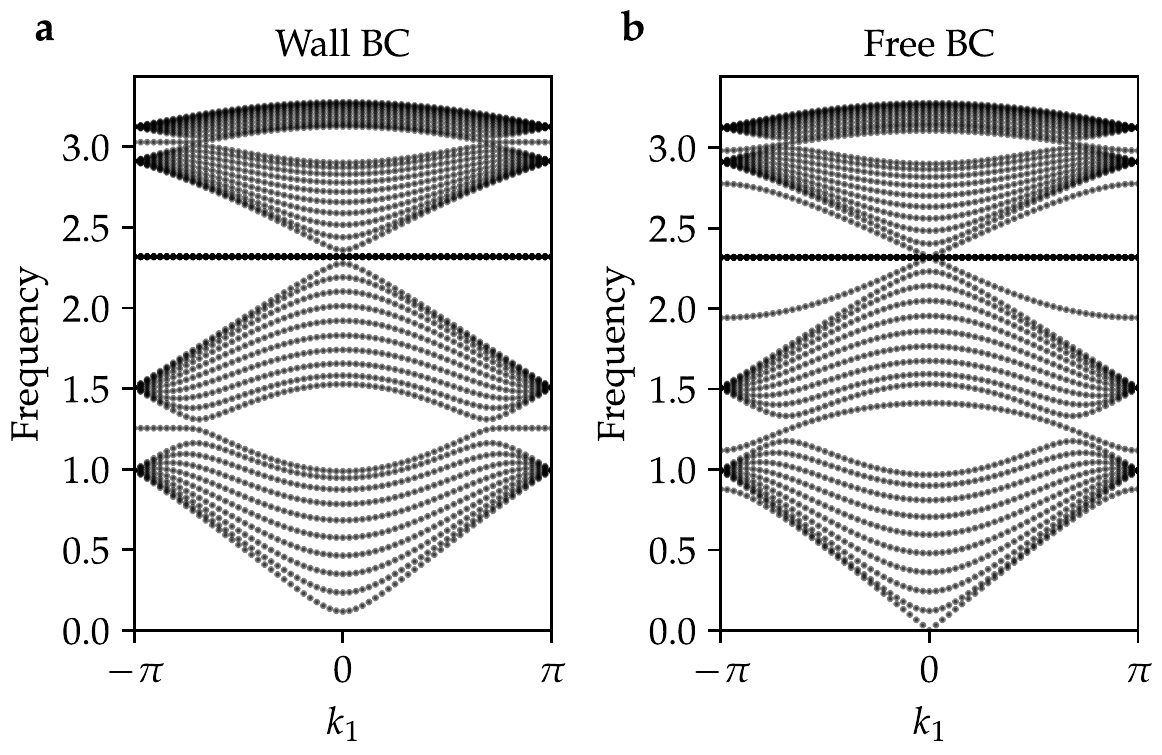}
    \caption{Edge states of the ribbon of honeycomb-kagome mass-spring model introduced in figure \ref{fig:honeycomb-kagome-ribbon} with $2m_{h}=3m_{k}$ for (a) the ideal `wall' boundary condition on the kagome sites (infinitely heavy capping masses at the honeycomb sites) and (b) the free boundary condition on the kagome sites (no capping masses). The edge states in (b) are not pinned to a particular energy because the honeycomb sector of the squared equations of motion is not chiral symmetric.}
    \label{fig:free-vs-wall-BC}
\end{figure}

We show in figure \ref{fig:SSH-void-channel-broken-chiral-symmetry}(a) the same SSH-like chain, as in figure \ref{fig:void-channel-SSH}, but without the capping voids. Figure \ref{fig:SSH-void-channel-broken-chiral-symmetry}(b) shows the squared frequency spectrum of the void-channel model (red crosses) and the corresponding mass-spring model with free boundary conditions (black points). As expected, there are no edge states because the chiral symmetry is strongly broken at the ends of the chains. Similarly, we verify in figure \ref{fig:free-vs-wall-BC} that the topological edge states are not present in the ribbon of the honeycomb-kagome lattice without heavy capping masses/large capping voids.

\section{Chiral symmetry of the squared honeycomb-kagome lattice for $m_h \neq m_k$}\label{app:chiral-sym-of-squared}

\begin{figure}
    \centering
    \includegraphics{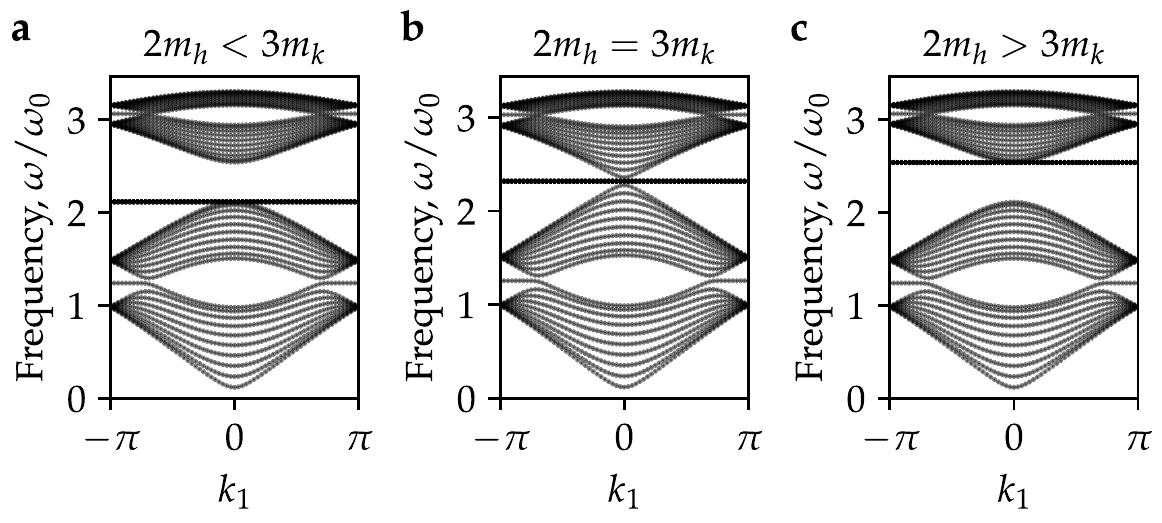}
    \caption{The edge states of the square-root semimetal are protected by the chiral symmetry of the squared equations, and may survive even as the chiral symmetry of the unsquared equations is broken. We plot the frequencies (normalised by $\omega=\sqrt{k_{0}/m_{0}}$) of the mass-spring model of the mass-spring ribbon from figure \ref{fig:honeycomb-kagome-ribbon} for (a) $2m_{h}<3m_{k}$, (b) $2m_{h}=3m_{k}$, and (c) $2m_{h}>3m_{k}$, where $m_{h}$ and $m_{k}$ are the masses at the honeycomb and kagome sites, respectively. The edge states are robust in all three systems even though the unsquared equations are not chiral symmetric for $2m_{h}\neq3m_{h}$. This is because the topological edge states are protected by the chiral symmetry of the honeycomb sector of the squared equations \cite{mizoguchi2021square}, which can be preserved even as the chiral symmetry of the unsquared equations is lost.}
    \label{fig:edge-states-vary-mh-mk}
\end{figure}

In figure \ref{fig:edge-states-vary-mh-mk} we plot the frequency bands of the mass-spring honeycomb-kagome network with perfect wall boundary conditions, but relax the constraint $2m_h=3m_k$ such that $\beta\neq0$ where $\beta$ is defined in equation \eqref{eq:beta}). This breaks the chiral symmetry of the unsquared equations, yet the edge states remain flat and robust against this perturbation, because the honeycomb sector of the \emph{squared} equations remains chiral symmetric.

\clearpage

\section*{References}

\bibliographystyle{unsrt}
\bibliography{references}

\begin{thebibliography}{10}

\bibitem{kosterlitz1973ordering}
J~M Kosterlitz and D~J Thouless.
\newblock Ordering, metastability and phase transitions in two-dimensional
  systems.
\newblock {\em Journal of Physics C: Solid State Physics}, 6(7):1181, 1973.

\bibitem{haldane1983continuum}
F~D~M Haldane.
\newblock Continuum dynamics of the 1-d heisenberg antiferromagnet:
  Identification with the $o(3)$ nonlinear sigma model.
\newblock {\em Physics Letters A}, 93(9):464--468, 1983.

\bibitem{haldane1983nonlinear}
F~D~M Haldane.
\newblock Nonlinear field theory of large-spin heisenberg antiferromagnets:
  semiclassically quantized solitons of the one-dimensional easy-axis n{\'e}el
  state.
\newblock {\em Physical Review Letters}, 50(15):1153, 1983.

\bibitem{checkelsky2012dirac}
J~G Checkelsky, J~Ye, Y~Onose, Y~Iwasa, and Y~Tokura.
\newblock Dirac-fermion-mediated ferromagnetism in a topological insulator.
\newblock {\em Nature Physics}, 8(10):729--733, 2012.

\bibitem{chiu2016classification}
C-K Chiu, J~C~Y Teo, A~P Schnyder, and S~Ryu.
\newblock Classification of topological quantum matter with symmetries.
\newblock {\em Reviews of Modern Physics}, 88(3):035005, 2016.

\bibitem{burkov2010spin}
A~A Burkov and D~G Hawthorn.
\newblock Spin and charge transport on the surface of a topological insulator.
\newblock {\em Physical Review Letters}, 105(6):066802, 2010.

\bibitem{vali2015scheme}
M~Vali, D~Dideban, and N~Moezi.
\newblock A scheme for a topological insulator field effect transistor.
\newblock {\em Physica E: Low-dimensional Systems and Nanostructures},
  69:360--363, 2015.

\bibitem{he2019topological}
M~He, H~Sun, and Q~L He.
\newblock Topological insulator: {Spintronics} and quantum computations.
\newblock {\em Frontiers of Physics}, 14(4):43401, 2019.

\bibitem{zhang2018topological}
X~Zhang, M~Xiao, Y~Cheng, M-H Lu, and J~Christensen.
\newblock Topological sound.
\newblock {\em Communications Physics}, 1(1):1--13, 2018.

\bibitem{ozawa2019topological}
T~Ozawa, H~M Price, A~Amo, N~Goldman, M~Hafezi, L~Lu, M~C Rechtsman,
  D~Schuster, J~Simon, O~Zilberberg, et~al.
\newblock Topological photonics.
\newblock {\em Reviews of Modern Physics}, 91(1):015006, 2019.

\bibitem{von2020fourty}
K~von Klitzing, T~Chakraborty, P~Kim, V~Madhavan, X~Dai, J~McIver, Y~Tokura,
  L~Savary, D~Smirnova, A~M Rey, et~al.
\newblock 40 years of the quantum {Hall} effect.
\newblock {\em Nature Reviews Physics}, pages 1--5, 2020.

\bibitem{kim2020recent}
M~Kim, Z~Jacob, and J~Rho.
\newblock Recent advances in 2d, 3d and higher-order topological photonics.
\newblock {\em Light: Science \& Applications}, 9(1):1--30, 2020.

\bibitem{rider2019perspective}
M~S Rider, S~J Palmer, S~R Pocock, X~Xiao, P~Arroyo~Huidobro, and V~Giannini.
\newblock A perspective on topological nanophotonics: current status and future
  challenges.
\newblock {\em Journal of Applied Physics}, 125(12):120901, 2019.

\bibitem{bandres2018topological}
M~A Bandres, S~Wittek, G~Harari, M~Parto, J~Ren, M~Segev, D~N Christodoulides,
  and M~Khajavikhan.
\newblock Topological insulator laser: {Experiments}.
\newblock {\em Science}, 359(6381), 2018.

\bibitem{ota2018topological}
Y~Ota, R~Katsumi, K~Watanabe, S~Iwamoto, and Y~Arakawa.
\newblock Topological photonic crystal nanocavity laser.
\newblock {\em Communications Physics}, 1(1):1--8, 2018.

\bibitem{khanikaev2013photonic}
A~B Khanikaev, S~H Mousavi, W-K Tse, M~Kargarian, A~H MacDonald, and G~Shvets.
\newblock Photonic topological insulators.
\newblock {\em Nature Materials}, 12(3):233--239, 2013.

\bibitem{schnyder2008classification}
A~P Schnyder, S~Ryu, A~Furusaki, and A~W~W Ludwig.
\newblock Classification of topological insulators and superconductors in three
  spatial dimensions.
\newblock {\em Physical Review B}, 78(19):195125, 2008.

\bibitem{kitaev2009periodic}
A~Kitaev.
\newblock Periodic table for topological insulators and superconductors.
\newblock In {\em AIP conference proceedings}, volume 1134, pages 22--30.
  American Institute of Physics, 2009.

\bibitem{ryu2010topological}
S~Ryu, A~P Schnyder, A~Furusaki, and A~W~W Ludwig.
\newblock Topological insulators and superconductors: tenfold way and
  dimensional hierarchy.
\newblock {\em New Journal of Physics}, 12(6):065010, 2010.

\bibitem{asboth2016short}
J~K Asb{\'o}th, L~Oroszl{\'a}ny, and A~P{\'a}lyi.
\newblock A short course on topological insulators.
\newblock {\em Lecture notes in physics}, 919:997--1000, 2016.

\bibitem{arkinstall2017topological}
J~Arkinstall, M~H Teimourpour, L~Feng, R~El-Ganainy, and H~Schomerus.
\newblock Topological tight-binding models from nontrivial square roots.
\newblock {\em Physical Review B}, 95(16):165109, 2017.

\bibitem{pocock2019bulk}
S~R Pocock, P~A Huidobro, and V~Giannini.
\newblock Bulk-edge correspondence and long-range hopping in the topological
  plasmonic chain.
\newblock {\em Nanophotonics}, 8(8):1337--1347, 2019.

\bibitem{pocock2020thesis}
S~Pocock.
\newblock {\em Topological physics in one-dimensional chains of metallic
  nanoparticles}.
\newblock PhD thesis, Imperial College London, 2020.

\bibitem{poli2015selective}
C~Poli, M~Bellec, U~Kuhl, F~Mortessagne, and H~Schomerus.
\newblock Selective enhancement of topologically induced interface states in a
  dielectric resonator chain.
\newblock {\em Nature Communications}, 6(1):1--5, 2015.

\bibitem{malkova2009observation}
N~Malkova, I~Hromada, X~Wang, G~Bryant, and Z~Chen.
\newblock Observation of optical {S}hockley-like surface states in photonic
  superlattices.
\newblock {\em Optics Letters}, 34(11):1633--1635, 2009.

\bibitem{vanel2017asymptotic}
A~L Vanel, O~Schnitzer, and R~V Craster.
\newblock Asymptotic network models of subwavelength metamaterials formed by
  closely packed photonic and phononic crystals.
\newblock {\em EPL (Europhysics Letters)}, 119(6):64002, 2017.

\bibitem{vanel2018asymptotic}
A~Vanel.
\newblock {\em Asymptotic analysis of discrete and continuous periodic media}.
\newblock PhD thesis, Imperial College London, 2018.

\bibitem{maimaiti2020microwave}
W~Maimaiti, B~Dietz, and A~Andreanov.
\newblock Microwave photonic crystals as an experimental realization of a
  combined honeycomb-kagome lattice.
\newblock {\em Physical Review B}, 102(21):214301, 2020.

\bibitem{wakao2020topological}
H~Wakao, T~Yoshida, T~Mizoguchi, and Y~Hatsugai.
\newblock Topological modes protected by chiral and two-fold rotational
  symmetry in a spring-mass model with a {L}ieb lattice structure.
\newblock {\em Journal of the Physical Society of Japan}, 89(8):083702, 2020.

\bibitem{Zheng_2019}
Li-Yang Zheng, Vassos Achilleos, Olivier Richoux, Georgios Theocharis, and
  Vincent Pagneux.
\newblock Observation of edge waves in a two-dimensional su-schrieffer-heeger
  acoustic network.
\newblock {\em Phys. Rev. Applied}, 12:034014, Sep 2019.

\bibitem{Zheng_2020}
L-Y Zheng, V~Achilleos, Z-G Chen, O~Richoux, G~Theocharis, Y~Wu, J~Mei,
  S~Felix, V~Tournat, and V~Pagneux.
\newblock Acoustic graphene network loaded with {H}elmholtz resonators: a
  first-principle modeling, {D}irac cones, edge and interface waves.
\newblock {\em New Journal of Physics}, 22(1):013029, jan 2020.

\bibitem{ZHENG2021100299}
L-Y Zheng, X-J Zhang, M-H Lu, Y-F Chen, and J~Christensen.
\newblock Knitting topological bands in artificial sonic semimetals.
\newblock {\em Materials Today Physics}, 16:100299, 2021.

\bibitem{vanel2019asymptotic}
A~L Vanel, O~Schnitzer, and R~V Craster.
\newblock Asymptotic modeling of phononic box crystals.
\newblock {\em SIAM J. Appl. Math.}, 79(2):506--524, 2019.

\bibitem{sun2012}
Y.~Sun, B.~Edwards, A.~Alu, and N.~Engheta.
\newblock Experimental realization of optical lumped nanocircuits at infrared
  wavelengths.
\newblock {\em Nat. Mater.}, 11:208--212, 2012.

\bibitem{matlack_designing_2018}
K~H Matlack, M~Serra-Garcia, A~Palermo, S~D Huber, and C~Daraio.
\newblock Designing perturbative metamaterials from discrete models.
\newblock {\em Nature Materials}, 17(4):323--328, 2018.

\bibitem{FreeFEM}
F.~Hecht.
\newblock New development in {FreeFem++}.
\newblock {\em Journal of Numerical Mathematics}, 20(3-4):251--265, 2012.

\bibitem{laude2015phononic}
V~Laude.
\newblock {\em Phononic crystals: artificial crystals for sonic, acoustic, and
  elastic waves}.
\newblock Number Vol. 26 in De {Gruyter} studies in mathematical physics. De
  Gruyter, Berlin, 2015.

\bibitem{su1980soliton}
W~P Su, J~R Schrieffer, and A~J Heeger.
\newblock Soliton excitations in polyacetylene.
\newblock {\em Physical Review B}, 22(4):2099, 1980.

\bibitem{ni2019observation}
X~Ni, M~Weiner, A~Alu, and A~B Khanikaev.
\newblock Observation of higher-order topological acoustic states protected by
  generalized chiral symmetry.
\newblock {\em Nature Materials}, 18(2):113--120, 2019.

\bibitem{mizoguchi2021square}
T~Mizoguchi, T~Yoshida, and Y~Hatsugai.
\newblock Square-root topological semimetals.
\newblock {\em Physical Review B}, 103(4):045136, 2021.

\bibitem{schnyder2018lecture}
A~P Schnyder.
\newblock Lecture notes on accidental and symmetry-enforced band crossings in
  topological semimetals.
\newblock {\em Topological Matter School, San Sebastian, Spain}, 2018.

\bibitem{barreteau2017bird}
C~Barreteau, F~Ducastelle, and T~Mallah.
\newblock A bird’s eye view on the flat and conic band world of the honeycomb
  and {K}agome lattices: towards an understanding of {2D} metal-organic
  frameworks electronic structure.
\newblock {\em Journal of Physics: Condensed Matter}, 29(46):465302, 2017.

\bibitem{delplace2011zak}
P~Delplace, D~Ullmo, and G~Montambaux.
\newblock Zak phase and the existence of edge states in graphene.
\newblock {\em Physical Review B}, 84(19):195452, 2011.

\bibitem{yan2020acoustic}
M~Yan, X~Huang, L~Luo, J~Lu, W~Deng, and Z~Liu.
\newblock Acoustic square-root topological states.
\newblock {\em Physical Review B}, 102(18):180102, 2020.

\end{thebibliography}

\end{document}